\documentclass[AMA,STIX1COL]{WileyNJD-v2}
\usepackage{geometry}
\usepackage{graphicx}
\usepackage{subfig}
\usepackage{caption}
\usepackage{makecell} 
\usepackage{float} 
 
\articletype{Article Type}%

\graphicspath{{./figures/}}
\begin{document}

\title{A Parallel Scalable Domain Decomposition Preconditioner for Elastic Crack Simulation Using XFEM}

\author[1,2]{Wei Tian}

\author[3]{Huang Jingjing}

\author[1,2]{Rongliang Chen*}

\author[1,2]{Yi Jiang}

\authormark{Wei Tian \textsc{et al}}

\address[1]{\orgdiv{}, \orgname{Shenzhen Key Laboratory for Exascale Engineering and Scientific Computing}, \orgaddress{\state{Shenzhen 518055}, \country{P.R. China}}}

\address[2]{\orgdiv{Shenzhen Institutes of Advanced Technology}, \orgname{Chinese Academy of Sciences}, \orgaddress{\state{Shenzhen 518055}, \country{P.R. China}}}

\address[3]{\orgdiv{Department of Rehabilitation Medicine}, \orgname{Shenzhen University General Hospital}, \orgaddress{\state{Shenzhen 518055},\country{China}}}

\corres{*Rongliang Chen, Shenzhen Institutes of Advanced Technology, Chinese Academy of Sciences, Shenzhen, P.R. China. \email{rl.chen@siat.ac.cn}}


\abstract[Abstract]{In this paper, a parallel overlapping domain decomposition preconditioner is proposed to solve the linear system of equations arising from the extended finite element discretization of elastic crack problems. The algorithm partitions the computational mesh into two types of subdomains: the regular subdomains and the crack tip subdomains based on the observation that the crack tips have a significant impact on the convergence of the iterative method while the impact of the crack lines is not that different from those of regular mesh points. The tip subdomains consist of mesh points at crack tips and all neighboring points where the branch enrichment functions are applied. The regular subdomains consist of all other mesh points, including those on the crack lines. To overcome the mismatch between the number of subdomains and the number of processor cores, the proposed method is divided into two steps: solve the crack tip problem and then the regular subdomain problem during each iteration. The proposed method was used to develop a parallel XFEM package which is able to test different types of iterative methods. To achieve good parallel efficiency, additional methods were introduced to reduce communication and to maintain the load balance between processors. Numerical experiments indicate that the proposed method significantly reduces the number of iterations and the total computation time compared to the classical methods. In addition, the method scales up to 8192 processor cores with over 70\% parallel efficiency to solve problems with more than $2\times10^8$ degrees of freedom.}

\keywords{elastic crack, XFEM, preconditioner, domain decomposition method, parallel computing}

\jnlcitation{\cname{%
\author{Wei Tian},
\author{Rongliang Chen},
\author{Yi Jiang},and
\author{Xiao-Chuan Cai}} (\cyear{2021}),
\ctitle{A Parallel Domain Decomposition Preconditioner for Elastic Crack Simulation Using XFEM}, \cjournal{International Journal for Numerical Methods in Engineering}, \cvol{2021;00:1--6}.}

\maketitle


\section{Introduction}\label{sec1}

Crack simulation in elastic materials is one of the most challenging problems in engineering applications. The extended finite element method (XFEM), which is widely appied to such problems, extends the polynomial approximation spaces to include additional enriched degrees of freedom (DOFs) to capture the discontinuities and singularities\cite{ref1,ref2}. XFEM has been successfully used for many problems, such as the simulation of fatigue crack growth in materials with inclusions and voids \cite{ref3,ref4}, the simulation of multi-material fracture propagation \cite{ref5,ref6}, and the computation of immiscible multi-phase flows \cite{ref7,ref8}.

As a result of the strong singularity at the crack tip, the system of linear equations arising from the XFEM discretization for crack problems is highly ill-conditioned \cite{ref9,ref10}. The condition number of the matrix from the standard finite element method (FEM) grows in the order $O(h^{-2}$), while it is $O(h^{-4}$) or worse for XFEM \cite{ref11}, where $h$ denotes the mesh size. The discretized system is difficult to solve using standard iterative methods because of the ill-conditioned matrix, especially when the size of the problem is large. There are many techniques to reduce the singularities of crack problems. For example, Lang, Sharma, and coworkers suggested the use of the Heaviside enrichment function to capture the discontinuities\cite{ref12,ref13}, while Menouillard, Elguedj, and coworkers used part of the branch enrichment functions to represent the crack tip physics \cite{ref14,ref15}. These methods make the resulting system easier to solve, but the accuracy and order of convergence are reduced. Another approach is to replace some of the polynomial approximation spaces by special shape functions without any extra DOFs, called the intrinsic XFEM \cite{ref16,ref17,ref18} like the iXFEM \cite{ref19,ref20}. Although these methods require more complex shape functions and result in a larger stiffness matrix bandwidth, the size of the algebraic system is smaller than that of the standard XFEM. The other class of strategies is to preserve the discretization scheme but seek efficient preconditioning techniques to reduce the condition number of the discretized system. There are several methods, including the additive Schwarz preconditioner \cite{ref21,ref22} and the multigrid preconditioner \cite{ref25,ref26}. These methods treat the entire crack area as a special domain and do not distinguish the crack line and crack tip area where the singularities mainly come from. These methods have been shown to work well for a variety of problems, however their parallel performance still needs to be explored when the scale of the problem and the number of processor cores are large.

In this paper, we refer to the approach of Chen and Cai \cite{ref24} to study the parallel scalability of an efficient domain decomposition preconditioner for the elastic crack problem. The main innovation of the algorithm is based on the observation that the singularities of the algebraic system mainly originate from the crack tip area. The normal part from standard FEM and the enriched part from Heaviside enrichment do not result in a large condition number. Therefore, a straightforward idea is to separate the crack tip subdomain and handle it individually. In Chen's paper, the algorithm is verified using a simple crack model with a small mesh in the MATLAB platform. There are still remaining issues that need to be investigated, such as whether the method works for complex crack models with a large number of processors. Based on Chen and Cai's work, a parallel version of their method is proposed and developed into a parallel package. In parallel computing, the load balance should be maintained and communications between processors should be decreased. The enrichment operations around cracks result in a non-uniform distribution of DOFs between processors, and the crack tip subdomains in the algorithm make the number of subdomains larger than the number of processors. These problems in parallel computing have to be addressed to increase scalability and parallel efficiency. In addition, to ensure the convenience of implementation, the submatrix for each crack tip is aggregated into one processor and solved locally because the crack may propagate anywhere in the computational domain. Based on our numerical analysis, this approach is rational because the communication between processors for the aggregation operation can be ignored as the main time-consuming part is to solve the regular subdomain problem.

The remainder of this paper is organized as follows. The governing equations and domain decomposition preconditioner are presented in Section \ref{sec2}. A benchmark case is used to verify the parallel XFEM code in Section \ref{sec3}. The numerical results are presented and discussed in Section \ref{sec4}. Finally, concluding remarks are presented in Section \ref{sec5}.

\section{Methodology}\label{sec2} 

\subsection{The governing equations and discretization} \label{sec2.1}

\begin{figure}[htb] 
\centerline{\includegraphics[scale=0.15]{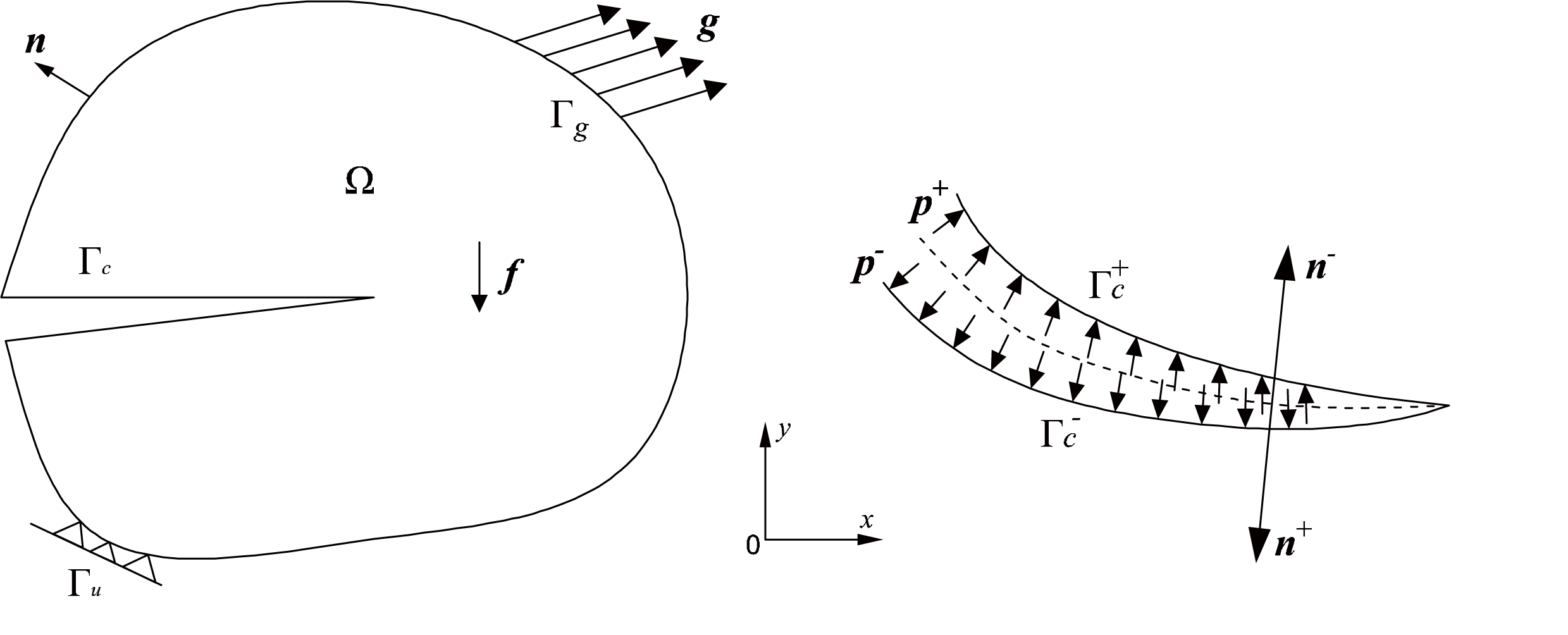}}
\caption{Diagram of the two-dimensional crack model (the dotted line is the crack line and is split into two crack surfaces, $\textit{\textbf{n}}^-$ and $\textit{\textbf{n}}^+$ are the unit normal vectors to the upper and bottom surfaces of the crack, respectively) \label{fig1}}
\end{figure}

In this study, we consider an elastostatic crack problem. As illustrated in Figure \ref{fig1}, the domain $\Omega\subset\mathbb{R}^2$ contains an edge crack $\Gamma_{\emph{c}}$, where the internal pressure along the crack surface is $p$. A prescribed displacement $\textbf{\emph{u}}_{0}$ is imposed on part of the boundary $\Gamma_{\emph{u}}$, and the rest of the boundary $\Gamma_{g}=\partial\Omega\backslash\Gamma_{u}$ is subject to a surface force \textbf{\emph{g}}. The unit normal to the boundaries is denoted as \textbf{\emph{n}}. The crack surface is distinguished by $\Gamma_{\emph{c}}^{+}$ and $\Gamma_{\emph{c}}^{-}$, such that $\Gamma_{\emph{c}}=\Gamma_{\emph{c}}^{+}\cup\Gamma_{\emph{c}}^{-}$ and with the unit normal to $\Gamma_{\emph{c}}^{+}$ and $\Gamma_{\emph{c}}^{-}$ by $\textbf{\emph{n}}^{+}$ and $\textbf{\emph{n}}^{-}$, respectively. We define $\textbf{\emph{n}}^{-}=-\textbf{\emph{n}}^{+}=\textbf{\emph{n}}$, and $\textit{\textbf{p}}^{+}=-\textit{\textbf{p}}^{-}=\textbf{\textit{p}}$, where $\textbf{\textit{p}}^{+}=-p\textit{\textbf{n}}^+$, $\textbf{\textit{p}}^{-}=-p\textbf{\textit{n}}^-$. The equilibrium equations and boundary conditions are given by \cite{ref28}
\begin{eqnarray} \label{eq1}
\left\{
\begin{aligned}
    \nabla\cdot\boldsymbol\sigma+\textbf{\emph{f}}&=0,            &\text{in}\ \Omega,\ \ \\
    \boldsymbol\sigma\cdot\textbf{\emph{n}}&=\textbf{\emph{g}},\  &\text{on}\ \Gamma_{g},\\
    \boldsymbol\sigma\cdot\textbf{\emph{n}}&=\textbf{\textit{p}},                   &\text{on}\ \Gamma_{c},\\
    \textbf{\emph{u}}&=\textbf{\emph{u}}_{0},          &\text{on}\ \Gamma_{u},\\
\end{aligned}
\right.
\end{eqnarray}
where $\emph{\textbf{f}}$ is a given body force, $\boldsymbol{\sigma}$ is the stress tensor corresponding to the displacement field $\emph{\textbf{u}}$, and $\nabla$ is the gradient operator $\nabla=\left(\partial/\partial x, \partial/\partial y\right)$. For the linear elasticity problem, the relationship between the stress $\boldsymbol{\sigma}$ and strain $\boldsymbol{\varepsilon}$ is
\begin{eqnarray} \label{eq2}
	\boldsymbol{\sigma}=\lambda\text{trace}(\boldsymbol{\varepsilon})\textbf{\textit{I}}+2\mu\boldsymbol{\varepsilon},\
	\boldsymbol{\varepsilon}=\frac{1}{2}(\nabla\emph{\textbf{u}}+(\nabla\emph{\textbf{u}})^{T}),
\end{eqnarray}
where "trace" is the trace of a tensor, and $\textbf{\emph{I}}$ is the identity matrix. The material Lam\'{e} constants $\mu$ and $\lambda$ are functions of the Young's modulus $\emph{E}$ and Poisson's ratio $\emph{v}$,
\begin{eqnarray} \label{eq3}
\mu=\frac{\emph{E}}{2(1+\nu)}\quad {\rm and}\quad \lambda=
\left\{
\begin{aligned}
    \frac{\emph{E}\nu}{(1+\nu)(1-2\nu)}\qquad &\rm for\ plain\ strain,\\
    \frac{\emph{E}\nu}{1-\nu^{2}}\qquad\qquad&\rm for\ plain\ stress.\\
\end{aligned}
\right.
\end{eqnarray}

The solution of (\ref{eq1}) lies in the space of admissible displacement field
\begin{eqnarray} \label{eq4}
\textbf{\textit{U}}=\{\textbf{\textit{u}}\,|\,\textit{\textbf{u}}\in [\textbf{\textit{H}}^{1}(\Omega\backslash\Gamma_{c})]^2,\textit{\textbf{u}}=\textit{\textbf{u}}_{0}\ \text {on}\ \Gamma_{u},\ \textit{\textbf{u}}\ \text {is discontinous across}\ \Gamma_{c}\},
\end{eqnarray}
where $\emph{\textbf{H}}^{1}(*)$ is the usual Sobolev space related to the regularity of the solution. A detailed description of the problem in a nonsmooth domain can be found in the work of Grisvard \cite{ref44}. Similarly, the test function space can be defined as
\begin{eqnarray} \label{eq5}
\textbf{\textit{V}}=\{\textit{\textbf{v}}\,|\,\textit{\textbf{v}}\in [\textbf{\textit{H}}^{1}_0(\Omega\backslash\Gamma_{c})]^2,\textit{\textbf{v}}=0\ \text {on}\ \Gamma_{u},\ \  \textit{\textbf{v}}\ \text{is discontinous across}\ \Gamma_{c}\}.
\end{eqnarray}

The variational formulation of (\ref{eq1}) is to find $\textbf{\emph{u}}\in\emph{\textbf{U}}$ such that
\begin{eqnarray} \label{eq6}
a(\boldsymbol{u},\boldsymbol{v})=l(\boldsymbol{v}),\quad \forall\boldsymbol{v}\in\textit{\textbf{V}},
\end{eqnarray}
where the bilinear form \emph{a}($\cdot$,$\cdot$) and linear form \emph{l}($\cdot$) are defined as
\begin{eqnarray} \label{eq7}
\left\{
\begin{aligned}
a(\boldsymbol{u},\boldsymbol{v})&=\int_{\Omega}\nabla\boldsymbol{u}:\nabla\boldsymbol{v}d\Omega,    \\
l(\boldsymbol{v})&=\int_{\Omega}\boldsymbol{f}\cdot\boldsymbol{v}d\Omega+
\int_{\Gamma_{g}}\boldsymbol{g}\cdot\boldsymbol{v}d\Gamma+
\int_{\Gamma_{c}}\boldsymbol{p}\cdot\boldsymbol{v}d\Gamma.
\end{aligned}
\right.
\end{eqnarray}

\begin{figure}[htb] 
\centerline{\includegraphics[scale=0.12]{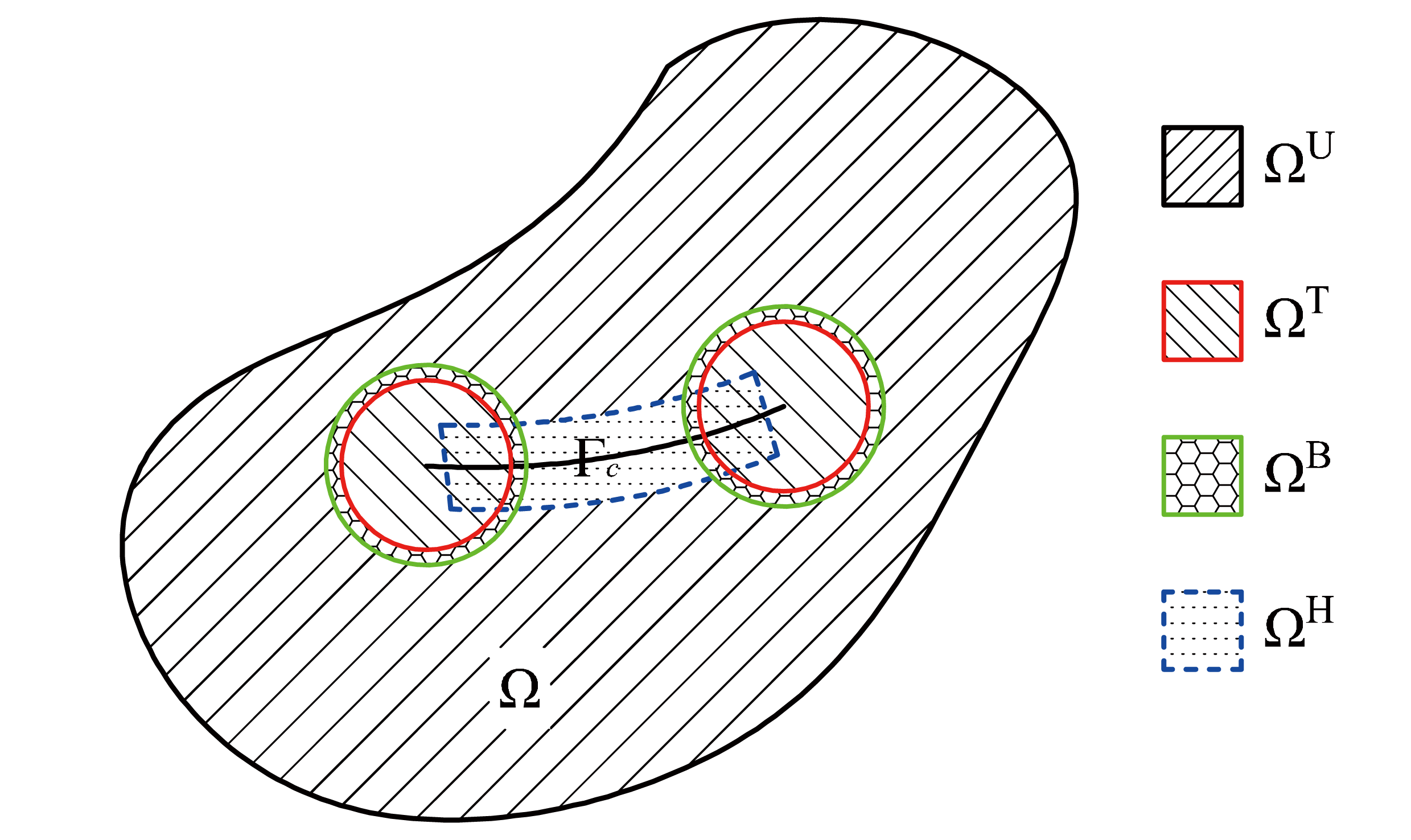}}
\caption{Diagram for the enriched domains and unenriched domain \label{fig2}}
\end{figure}

To discretize (\ref{eq6}), we apply the corrected XFEM\cite{ref29}, which uses a level set function to track the position of cracks, the branch enrichment functions to describe the asymptotic field around the crack tips, and the ramp function to handle the blending area. In this study, the computational domain was discretized by a structured Cartesian mesh with quadrilateral elements. As illustrated in Figure \ref{fig2}, the computational domain is initially divided into the crack tip enriched domain $\Omega^{\rm T}$, the blending area $\Omega^{\rm B}$, the Heaviside enriched domain $\Omega^{\rm H}$, and the unenriched domain $\Omega^{\rm U}$. The Heaviside enriched domain $\Omega^{\rm H}$ is the area where all elements are cut by the crack line, excluding the elements that contain the crack tip. The crack tip enriched domain $\Omega^{\rm T}$ is the area within a radius $r_{tip}$ of the crack tip, where all the nodes for each element are enriched by the branch enrichment functions. The blending domain $\Omega^{\rm B}$ is the area around $\Omega^{\rm T}$, where only a part of the nodes for each element are enriched by the branch enrichment functions. The unenriched domain $\Omega^{\rm U}$ is the area without any enriched nodes, such that $\Omega^{\rm U}=\Omega\backslash(\Omega^{\rm T}\cup\Omega^{\rm B}\cup\Omega^{\rm H})$. Note that the Heaviside enriched domain $\Omega^H$ may overlap with $\Omega^T$ and $\Omega^B$, and the nodes in the overlap area are enriched by multiple enrichment functions.

For elements in $\Omega^{\rm U}$, the displacement field is continuous, so the piecewise linear continuous function from the standard FEM can be adopted as the basis function. In the area around cracks, additional basis functions should be adopted to represent the physics of the problem, except for the standard basis function. For the elements in $\Omega^{\rm H}$, the variables cross cracks are discontinuous, and the additional basis function is the Heaviside function, which represents the discontinuity
\begin{eqnarray} \label{eq8}
H(x)=\left\{
\begin{aligned}
+1\qquad&\text{if}\ (\boldsymbol{x}-\boldsymbol{x}_{c})\cdot\textbf{\emph{n}}^{-}_{c}\geq0,\\
-1\qquad&\text{otherwise},
\end{aligned}
\right.
\end{eqnarray}
where $\textbf{\emph{n}}^{-}_{c}$ is the unit normal to the negative side of the crack surface, and $\textbf{\emph{x}}_{c}$ is a point on the crack surface that has the shortest distance to $\textbf{\emph{x}}$. For the elements in $\Omega^{\rm T}$, the analytical displacement fields for the combined mode I and II loading can be found in \cite{ref30}. The analytical solutions are not only discontinuous but also includes singularities. The branch enrichment functions can be selected from the basic terms of the analytical solutions.
\begin{eqnarray} \label{eq9}
\phi^{\alpha}(r(x), \theta(x))=\sqrt{r}\left\{\sin \left(\frac{\theta}{2}\right), \cos \left(\frac{\theta}{2}\right), \sin \left(\frac{\theta}{2}\right) \sin \theta, \cos \left(\frac{\theta}{2}\right) \sin \theta\right\},
\end{eqnarray}
where $\left(r,\theta\right)$ is the local polar coordinate corresponding to the crack tip. The approximation of the displacement field for the corrected XFEM can be formulated as follows \cite{ref29,ref31}
\begin{eqnarray} \label{eq10}
\begin{aligned}
\textbf{\textit{u}}_{h}(x) &=\sum\limits_{i \in N_{S}}N_{i}(x) \textit{\textbf{u}}_{i}+\sum\limits_{j \in N_{H}} N_{j}(x)\left[H(x)-H\left(x_{j}\right)\right] \textbf{\textit{a}}_{j}+\sum\limits_{k \in N_{ti p}} \sum\limits_{\alpha=1}^{4} N_{k}(x)\left[\phi^{\alpha}(x)-\phi^{\alpha}\left(x_{k}\right)\right] \mathcal{R}(x) \textit{\textbf{b}}_{k}^{\alpha},
\end{aligned}
\end{eqnarray}
where $\emph{N}_{S}$, $\emph{N}_{H}$, and $\emph{N}_{tip}$ represent the set of all mesh nodes, the nodal subset for elements cut by cracks, and the nodal subset for elements around the crack tips within an enrichment radius $\emph{r}_{tip}$, respectively; $\emph{N}_i$, $\emph{N}_j$, and $\emph{N}_k$ are the standard FEM shape functions; $\emph{u}_i$ is the unknown associated with the node set $\emph{N}_s$; $\emph{\textbf{a}}_j$ and $\emph{\textbf{b}}_{\alpha}^{k}$ are the unknowns for the enriched nodal subset $\emph{N}_{H}$ and $\emph{N}_{tip}$; $\mathcal{R}(x)$ represents a ramp function that is employed to overcome the problem of blending elements in $\Omega^{\text{B}}$ \cite{ref32}. Based on the approximation in (\ref{eq10}), the discrete system for (\ref{eq6}) has the following formulation:
\begin{eqnarray} \label{eq11}
\textbf{\textit{Kd}}=\textbf{\textit{F}},
\end{eqnarray}
where $\textbf{\textit{F}}$ is the force vector, $\textbf{\textit{d}}=(\textit{\textbf{u}},\textit{\textbf{a}},\textit{\textbf{b}})$ is the unknown, and $\textbf{\textit{K}}=(k_{ij})$, known as the stiffness matrix, is an $n\times n$ symmetrical sparse matrix, whose structure is as follows:
\begin{equation} \label{eq12}
	\textit{\textbf{K}}=
	\begingroup\renewcommand*{\arraystretch}{1.5}
	\begin{bmatrix}
		\textit{\textbf{k}}^{uu}& & \textit{\textbf{k}}^{ua} & &\textit{\textbf{k}}^{ub} \\
		\textit{\textbf{k}}^{au}& & \textit{\textbf{k}}^{aa} & &\textit{\textbf{k}}^{ab} \\
		\textit{\textbf{k}}^{bu}& & \textit{\textbf{k}}^{ba} & &\textit{\textbf{k}}^{bb} \\
	\end{bmatrix},
	\endgroup
\end{equation}
where $\textit{\textbf{k}}^{uu}$ is the stiffness matrix related to the standard FEM basis, which has the same nonzero structure as the standard FEM; $\textit{\textbf{k}}^{aa}$ and $\textit{\textbf{k}}^{bb}$ are the stiffness matrices related to the enriched DOFs, which are denser than $\textit{\textbf{k}}^{uu}$; and the other parts are the coupling matrices for regular DOFs and enriched DOFs.

In addition, the linear dependency for the local enrichment in each crack tip should be carefully considered, as described in the work of Fries\cite{ref15}. The detailed numerical method for eliminating the linear dependency can be found in Appendix A. Unlike the standard FEM, the matrix $\textbf{\textit{K}}$ from the corrected XFEM is highly ill-conditioned, and the linear system is difficult to solve \cite{ref33}. An efficient method for solving the ill-conditioned system is to use the preconditioned Krylov subspace with a well-designed preconditioner. The main focus of this study is to design an efficient and scalable preconditioner to solve the large-scale linear system (\ref{eq11}).

\subsection{The additive Schwarz method and algorithmic framework} \label{sec2.2}
Because of the singularities of the asymptotic field around the crack tip, the linear system (\ref{eq11}) is difficult to solve using classical iterative methods, such as GMRES and CG, especially for large-scale problems in parallel computing. To accelerate the iterative method, preconditioning techniques are required. One of the most popular preconditioning methods in parallel computing is the additive Schwarz method \cite{ref34}. Solving the linear system by combining the left preconditioned Krylov subspace method with the additive Schwarz method (\ref{eq11}) is equivalent to solving the following preconditioned system:
\begin{eqnarray} \label{eq13}
\textbf{\textit{M}}_*^{-1}\textbf{\textit{K}}\textbf{\textit{d}}=\textbf{\textit{M}}_*^{-1}\textbf{\textit{F}}	
\end{eqnarray}
where $\textbf{\textit{M}}_*^{-1}$ is the additive Schwarz preconditioner.

The additive Schwarz method is a type of overlapping domain decomposition method. This can be defined in two ways. The first is the algebraic method, which obtains the subdomain matrices $\textbf{\emph{K}}_i (i=1, 2, ..., N)$ and overlaps algebraically based on the matrix $\textbf{\textit{K}}$ \cite{ref35}. The second method is the geometrical method, which begins by partitioning the physical domain $\Omega$ into nonoverlapping subdomains $\Omega_i^0$ (i=1, 2, …, N) and then extending each subdomain $\Omega_i^0$ into an overlapping subdomain $\Omega_i^{\delta}$ by adding $\delta$ layers of elements from the neighboring subdomains. For each subdomain, the submatrix $\textbf{\emph{K}}_i$ is obtained as follows:
\begin{eqnarray} \label{eq14}
	\textbf{\textit{K}}_i=\textbf{\textit{R}}_i^{\delta}\textbf{\textit{K}}(\textbf{\textit{R}}_i^{\delta})^T,\ i=1,2,\dots,N,
\end{eqnarray}
where \textit{N} is the number of subdomains, $\textbf{\textit{R}}_i^\delta$ is a restriction operator from the global domain $\Omega$ to the overlapping subdomain $\Omega_i^\delta$, which is an $n_i\times n$ rectangular matrix, where $n$ is the global matrix size and $n_i$ is the local matrix size. Only one component in each row of $\textbf{\textit{R}}_i^{\delta} $ is set to one, which is related to the DOF belonging to $\Omega_i^{\delta}$, and the other components are set to zero. $(\textbf{\textit{R}}_i^{\delta})^T$ is the extension operator from local to global, which is usually the transpose of the restriction operator. Depending on the method of treating the overlap, there are three types of additive Schwarz preconditioners: classical additive Schwarz preconditioner, restricted additive Schwarz preconditioner, and harmonic additive Schwarz preconditioner. In this study, we adopt the classical additive Schwarz preconditioner (ASM), which is defined as
\begin{eqnarray} \label{eq15}
	\textbf{\textit{M}}_\text{ASM}^{-1}&=(\textbf{\textit{R}}_1^\delta)^T\textbf{\textit{B}}_1^{-1}\textbf{\textit{R}}_1^\delta+\cdots+(\textbf{\textit{R}}_\textit{N}^\delta)^T\textbf{\textit{B}}_\textit{N}^{-1}\textbf{\textit{R}}_\textit{N}^\delta
\end{eqnarray}
where $\textbf{\textit{B}}_i^{-1}$ is the approximation of the inverse matrix of $\textbf{\textit{K}}_i$, which is usually computed by the incomplete Cholesky factorization (ICC).

The numerical results in Section \ref{sec4.2} indicate that the algebraic additive Schwarz preconditioner and the standard geometrical additive Schwarz preconditioner do not work well for the crack problems considered in this study. The differences in the properties of the subdomains with and without cracks, which can be quite extensive, must be considered when constructing the additive Schwarz preconditioner. In this paper, a geometrical additive Schwarz preconditioner is considered after carefully examining the properties of the crack tip subdomains.

\begin{figure}[htb]
	\centerline{\includegraphics[scale=0.20]{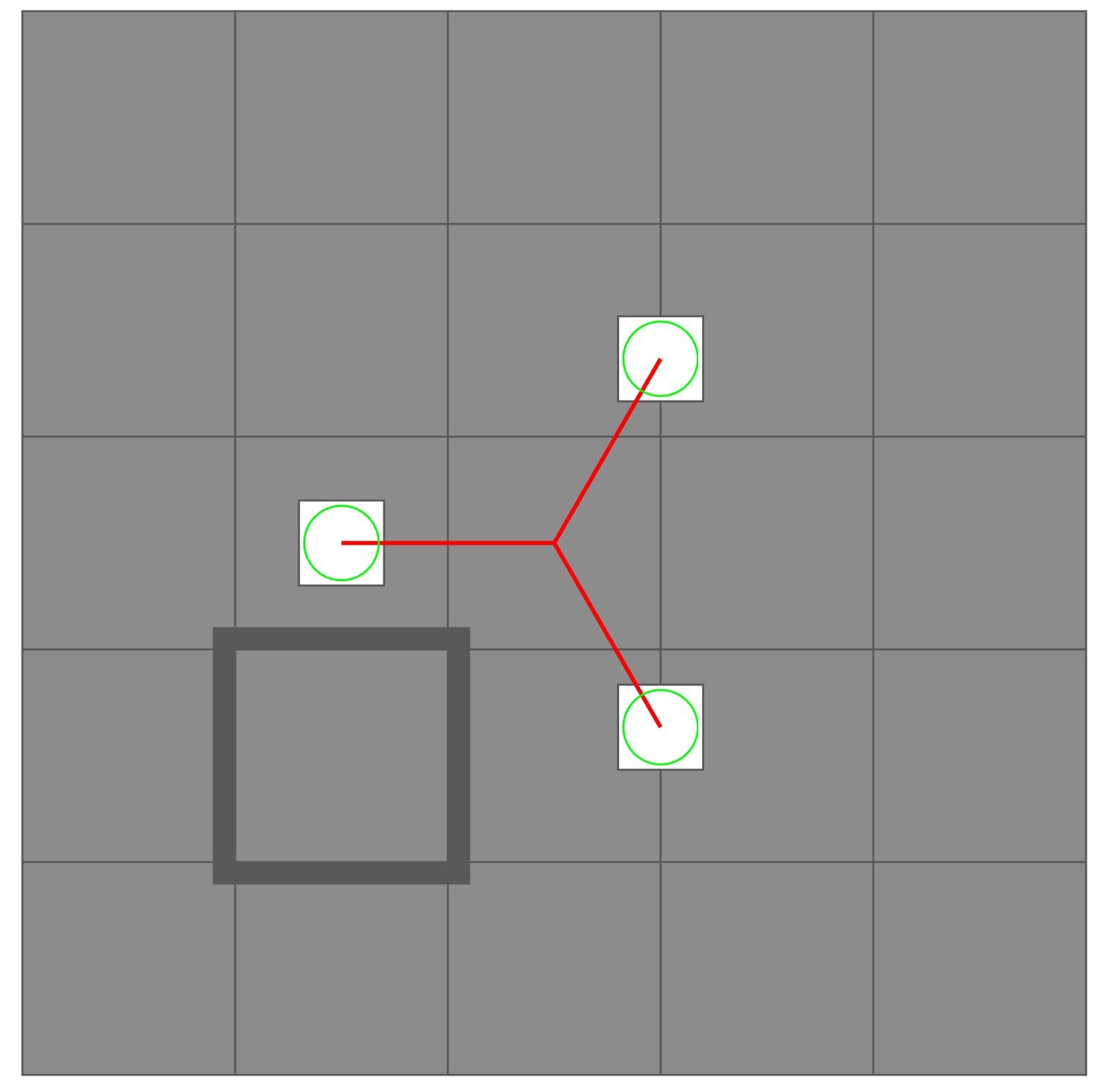}}
	\caption{Topology of subdomains for branch crack: the light gray represents regular subdomains; the white represents crack tip subdomains; the dark gray represents the subdomain overlap; the red line is crack line; and the green circle is the crack tip enrichment radius $r_{tip}$}
	\label{fig3}  
\end{figure}

As illustrated in Figure \ref{fig3}, all subdomains are classified into two types: regular subdomains $\Omega_i^r\ (i\in[1,N_{reg}])$ and crack tip subdomains $\Omega_j^t\ (j\in[1,N_{tip}])$. The regular subdomains are partitioned based on the geometry of the computational domain $\Omega$, regardless of the crack position. The number of regular subdomains $\textit{N}_{reg}$ is equal to the number of processors $\textit{N}_p$. The number of crack tip subdomains $\textit{N}_{tip}$ is equal to the number of crack tips. Thus, $\bigcup\Omega_j^t=\Omega^{\rm T}\cup\Omega^{\rm B}$ and $\bigcup\Omega_i^r=\Omega\backslash (\Omega^{\rm T} \cup \Omega^{\rm B})$. For parallel computing, the submatrices from regular subdomains are allocated to each processor locally, whereas the submatrices from the crack tip subdomains may be distributed over arbitrary processors. For convenience of calculation, the submatrices from the crack tip subdomains are aggregated and stored on specific $\textit{N}_{tip}$ processors to compute the inverse matrix of each subdomain according to (\ref{eq15}). To decrease the communication between processors, the $N_{tip}$ processors should include the enrichment area around the crack tips as much as possible.

During each iteration of the preconditioned Krylov subspace iterative method, we need to compute $\textbf{\textit{y}}=\textbf{\textit{M}}_*^{-1}\textbf{\textit{x}}$, where $\textbf{\textit{x}}$ is the residual vector or the vector of unknowns defined on the domain $\Omega$. The ASM preconditioner can be expressed as:
\begin{eqnarray} \label{eq16}
\textbf{\textit{y}}=\left[(\textbf{\textit{R}}_1^\delta)^T\textbf{\textit{B}}_1^{-1}\textbf{\textit{R}}_1^\delta+\cdots+(\textbf{\textit{R}}_\textit{N}^\delta)^T\textbf{\textit{B}}_\textit{N}^{-1}\textbf{\textit{R}}_\textit{N}^\delta\right] \textbf{\textit{x}}.
\end{eqnarray}
This was computed in parallel. That is, on the $i$th processor core, we compute
\begin{eqnarray} \label{eq17}
	\textbf{\textit{y}}_i=(\textbf{\textit{R}}_i^\delta)^T\textbf{\textit{B}}_i^{-1}\textbf{\textit{R}}_i^\delta\textbf{\textit{x}},
\end{eqnarray}
then obtain the global vector by adding them together
\begin{eqnarray} \label{eq18}
	\textbf{\textit{y}}=\sum_{i=1}^{N}\textbf{\textit{y}}_i,
\end{eqnarray}
where $\textbf{\textit{y}}_i$ includes two parts in the $N_{tip}$ processors, which include a regular subdomain and an enriched crack tip subdomain. The subvector in each processor core with overlap $\textbf{\textit{x}}_i^\delta$ is defined as
\begin{eqnarray} \label{eq19}
	\textbf{\textit{x}}_i^\delta=\textbf{\textit{R}}_i^\delta\textbf{\textit{x}}
\end{eqnarray}
Let $\textbf{\textit{X}}_i=\textbf{\textit{R}}_i^\delta\textbf{\textit{x}}=\textbf{\textit{x}}_i^\delta$ and $\textbf{\textit{Y}}_i=\textbf{\textit{B}}_i^{-1}\textbf{\textit{X}}_i$, then,  $\textbf{\textit{y}}_i=(\textbf{\textit{R}}_i^\delta)^T\textbf{\textit{Y}}_i$. $\textbf{\textit{Y}}_i$ is solved by solving the linear system of the equations $\textbf{\textit{K}}_i\textbf{\textit{Y}}_i=\textbf{\textit{X}}_i$ locally. For the specific $N_{tip}$ processors, we solve $\textbf{\textit{K}}_i^{reg}\textbf{\textit{Y}}_i^{reg}=\textbf{\textit{X}}_i^{reg}$ and $\textbf{\textit{K}}_i^{tip}\textbf{\textit{Y}}_i^{tip}=\textbf{\textit{X}}_i^{tip}$, and then $\textit{\textbf{y}}_i=(\textit{\textbf{R}}_{i,reg}^\delta)^T\textbf{\textit{Y}}_i^{reg}+(\textit{\textbf{R}}_{i,tip}^\delta)^T\textbf{\textit{Y}}_i^{tip}$. Here, we distinguish the restriction and extension operators by subscripts $reg$ and $tip$ in the $N_{tip}$ processors. This means that $(\textit{\textbf{R}}_{i,reg}^\delta)^T$ and $(\textit{\textbf{R}}_{i,tip}^\delta)^T$ are extension operators for regular subdomains and tip subdomains in the $N_{tip}$ processors. A subscript is not needed for the other $N_p-N_{tip}$ processors because only one subdomain problem needs to be solved by each processor. Note that the restriction and extension operators with subscript $tip$ deal with all types of DOFs, whereas operators with subscript $reg$ or without any subscript only deal with regular DOFs and Heaviside DOFs. The above algorithm is summarized Algorithm 1.
\begin{algorithm}[Ht!]  
\begin{enumerate}[\hspace{2em}(1)]
	\renewcommand{\labelenumi}{Step \theenumi:}
	\item Obtain the submatrices $\textbf{\textit{K}}_i^{tip}$ for tip subdomains and distribute them to the specific $N_{tip}$ processors.
	\item Obtain $\textit{\textbf{K}}_i^{reg}$ with overlaps in all processors.
	\item Each processor gets the subvector $\textbf{\textit{X}}_i$ from global vector $\textbf{\textit{x}}$.\\
	 For the specific $N_{tip}$ processors, $\textit{\textbf{X}}_i^{reg}=\textit{\textbf{R}}_{i,reg}^\delta \textbf{\textit{x}}$ and $\textit{\textbf{X}}_i^{tip}=\textit{\textbf{R}}_{i,tip}^\delta \textbf{\textit{x}}$.\\
	 For the other $N_p-N_{tip}$ processors, $\textit{\textbf{X}}_i^{reg}=\textit{\textbf{R}}_{i}^\delta \textbf{\textit{x}}$.
	\item In the specific $N_{tip}$ processors, solve the linear equations\\
	$\textbf{\textit{K}}_i^{tip}\textbf{\textit{Y}}_i^{tip}=\textbf{\textit{X}}_i^{tip} \left(i=1,2,...,N_{tip} \right)$,\\
	the other $N_p-N_{tip}$ processors do nothing and wait for these $N_{tip}$ processors.
	\item In all processors, solve the linear equations\\
	$\textbf{\textit{K}}_i^{reg}\textbf{\textit{Y}}_i^{reg}=\textbf{\textit{X}}_i^{reg} \left(i=1,2,...,N_{p} \right)$
	\item For the specific $N_{tip}$ processors, $\textit{\textbf{y}}_i=(\textit{\textbf{R}}_{i,reg}^\delta)^T\textbf{\textit{Y}}_i^{reg}+(\textit{\textbf{R}}_{i,tip}^\delta)^T\textbf{\textit{Y}}_i^{tip} \left(i=1,2,...,N_{tip} \right)$.\\
	For the other $N_p-N_{tip}$ processors, let $\textbf{\textit{y}}_i=(\textit{\textbf{R}}_{i}^\delta)^T\textbf{\textit{Y}}_i^{reg}\left(i=N_{tip}+1,N_{tip}+2,...,N_{p} \right)$.
	\item Assemble the vector \textbf{\textit{y}} by\\
	\begin{equation} \nonumber
		\textbf{\textit{y}}=\sum_{i=1}^{N_p}\textbf{\textit{y}}_i.
	\end{equation}
\end{enumerate}
	\caption{Additive Schwarz preconditioner based on geometrical domain decomposition}  
	\label{algo1}
\end{algorithm}

\begin{figure}[Htb]
	\centerline{\includegraphics[scale=0.60]{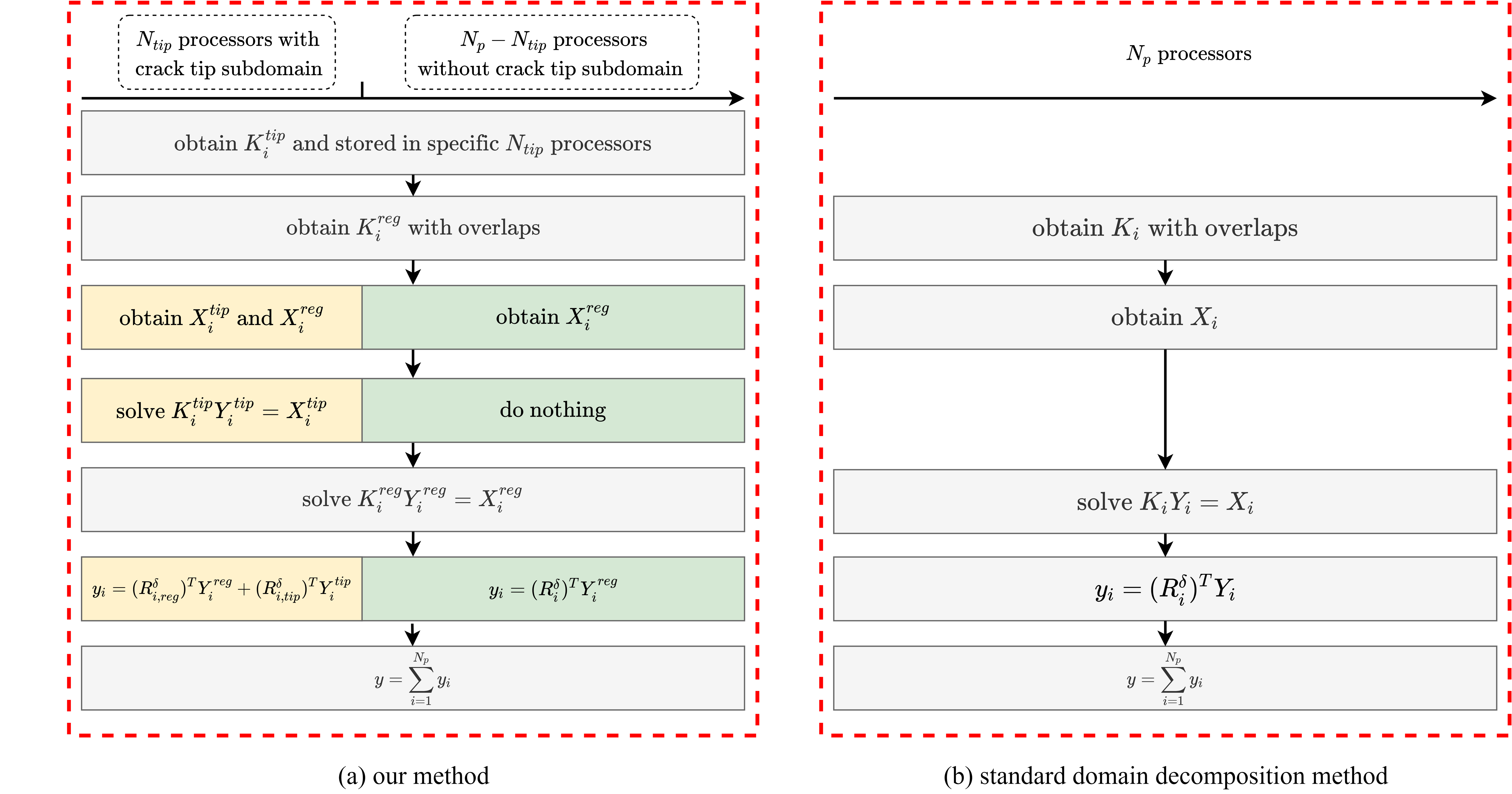}}
	\caption{The difference between our algorithm and standard domain decomposition method with additive Schwarz methdod} 
	\label{fig4}  
\end{figure}

The main idea of the algorithm is to handle the crack tip subdomain separately. The difference between this method and the standard domain decomposition method (sDDM) is depicted in Figure \ref{fig4}. In sDDM, all the subdomain problems are solved in parallel on each processor, and the communication between the processors is only from the overlaps. In our method, the crack tip submatrix is obtained and stored on specific $N_{tip}$ processors. To reduce communication, a flag in each processor indicates whether the processor has crack tips inside, and the data transfer mainly occurs among processors that include crack tips. If more than two processors share a crack tip area, the one that holds most of the crack tip occupies the subdomain and the problem is solved locally. During each iteration of the preconditioned Krylov subspace method, these $N_{tip}$ processors need to perform more work than sDDM because the linear system from the crack tip area needs to be solved while the other $N_p-N_{tip}$ processors are idle and wait for the $N_{tip}$ processors to finish the operation. This affects the synchronization between processors and hence the parallel efficiency. The numerical results in Section \ref{sec4.7} demonstrate that this effect is very small, as the most time-consuming step is to solve the linear equations from the regular subdomains. It takes very little time to solve the crack tip problem because the size of the crack tip submatrix is usually much smaller than that of the regular subdomain.  

\noindent\textbf{Remark 1}: In XFEM, the unknowns can be classified into three types: the normal DOFs from standard FEM, the Heaviside DOFs from elements cut by cracks, and the crack tip DOFs from the crack tip area. Accordingly, there are different strategies for choosing subdomain DOFs and overlapping DOFs in (\ref{eq14}). Here, three subdomain strategies are introduced referred to as the algebraic subdomain strategy S0, the regular geometrical subdomain strategy S1, and the advanced subdomain strategy S2. A comparison of their performances is discussed in Section \ref{sec4.2}.
\begin{enumerate}[\hspace{0.5em}S1:]
	\addtocounter{enumi}{-1}
	\setlength{\itemsep}{-0.8ex}
\item The subdomain DOFs and overlapping DOFs are determined based on the stiffness matrix $\textbf{\textit{K}}$ and totally ignore the geometry information.
\item The computational domain is divided into $N_p$ subdomains without consideration of the crack lines. Each subdomain include all types of DOFs. The overlap part for each subdomain has the same type of DOF as the inner part.
\item The computational domain is divided into $N_p$ regular subdomains and $N_{tip}$ crack tip subdomains. The regular subdomain includes normal DOFs and Heaviside DOFs, and the crack tip subdomain includes all types of DOFs. There are no interactions between the two types of subdomains. If a node is located in a crack tip subdomain, the corresponding DOFs no longer belong to a regular subdomain. The overlap for each subdomain has the same type of DOF as the corresponding subdomain.
\end{enumerate}

\noindent\textbf{Remark 2}: In practice, the inverse matrix $\textbf{\textit{M}}_*^{-1}$, which is usually dense and memory consuming, is not implemented. In the preconditioned Krylov subspace iterative method, the  explicit formulation of the entire preconditioner matrix is not required because the matrix-vector multiplications are only performed during each iteration. According to (\ref{eq15}), the preconditioner $\textbf{\textit{M}}_*^{-1}$ is a block diagonal matrix. When performing the matrix-vector multiplications, the multiplication of submatrices and subvectors is computed locally on each processor and then assembled into the global vector. For subdomain strategy S2, the number of subdomains is larger than the number of processors, and the crack tip subdomains have to be specially considered.
\section{Benchmark case}\label{sec3}
In this study, a parallel XFEM package on top of the open-source package PETSc \cite{ref45} was developed to study the above algorithm. To verify the accuracy and robustness of the code, a benchmark problem was employed to investigate the convergence rate, accuracy, and condition number of the coefficient matrix.

\begin{figure}[htb] 
	\centerline{\includegraphics[scale=0.2]{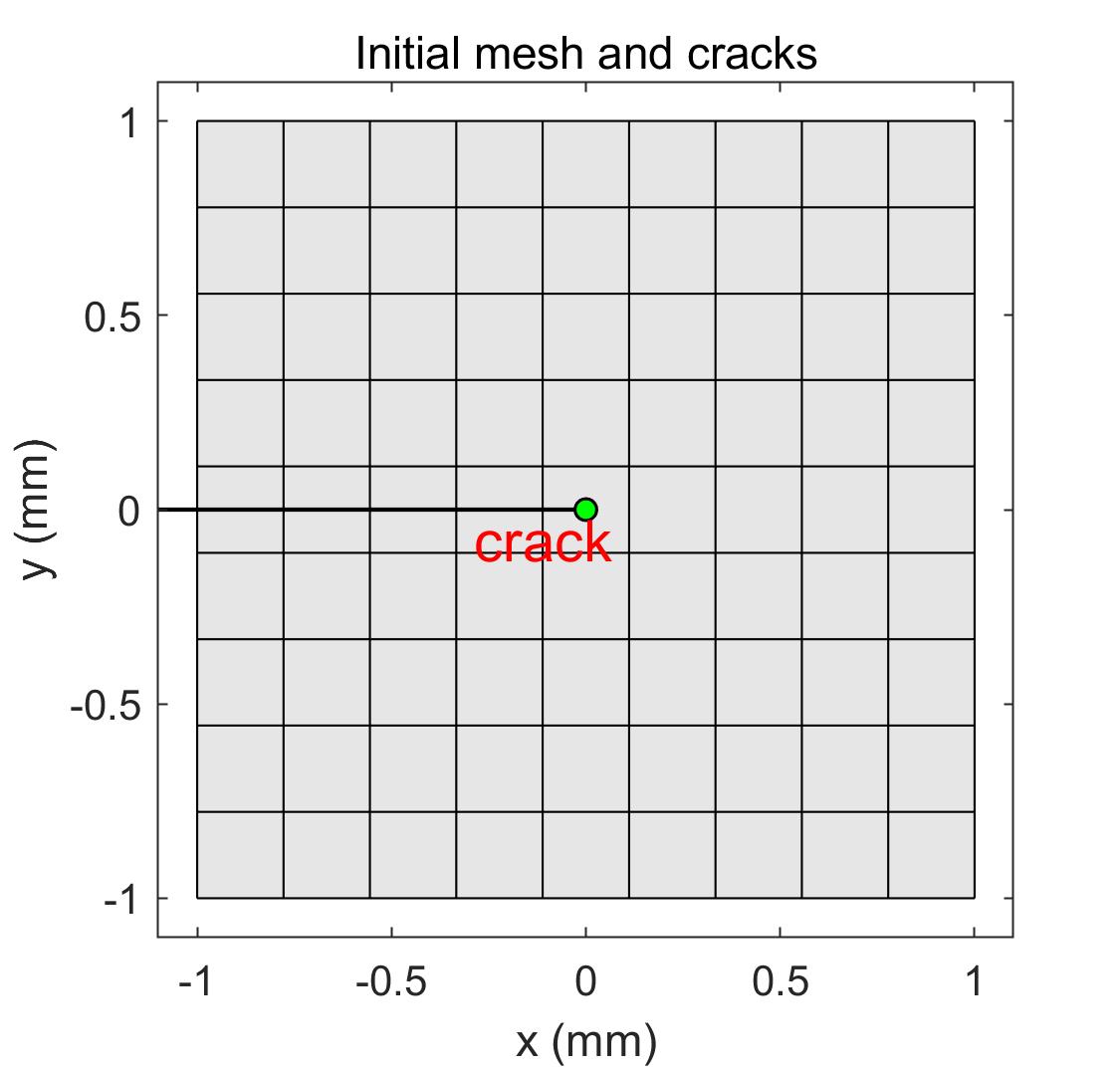}}
	\caption{Diagram of the edge crack model \label{fig5}}
\end{figure}

The benchmark problem is a two-dimensional edge-crack model, as shown in Figure \ref{fig5}. The domain is a square of $2m \times 2m$, and an edge crack is located in the left half of the domain. The problem type is plain stress. Young’s modulus $E=10^4$ MPa, and Poisson's ratio $\nu=0.30$. The mesh was refined from $10 \times 10$ to $400 \times 400$, and the corresponding mesh size changed from 0.222 to 0.005. The boundaries are subject to analytical displacement in the literature \cite{ref30}. For comparison, an FEM model without cracks was employed with the same material constants and mesh size. There are two strategies for the crack tip enrichment: topological enrichment (top-XFEM) and geometrical enrichment (geo-XFEM) \cite{ref36}. The top-XFEM method enriches the crack tip with fixed layers of elements, and the enrichment area varies with mesh refinement. The geo-XFEM method provides a fixed enrichment radius, regardless of the mesh refinement. 

\begin{figure}[htb] 
	\centerline{\includegraphics[scale=0.4]{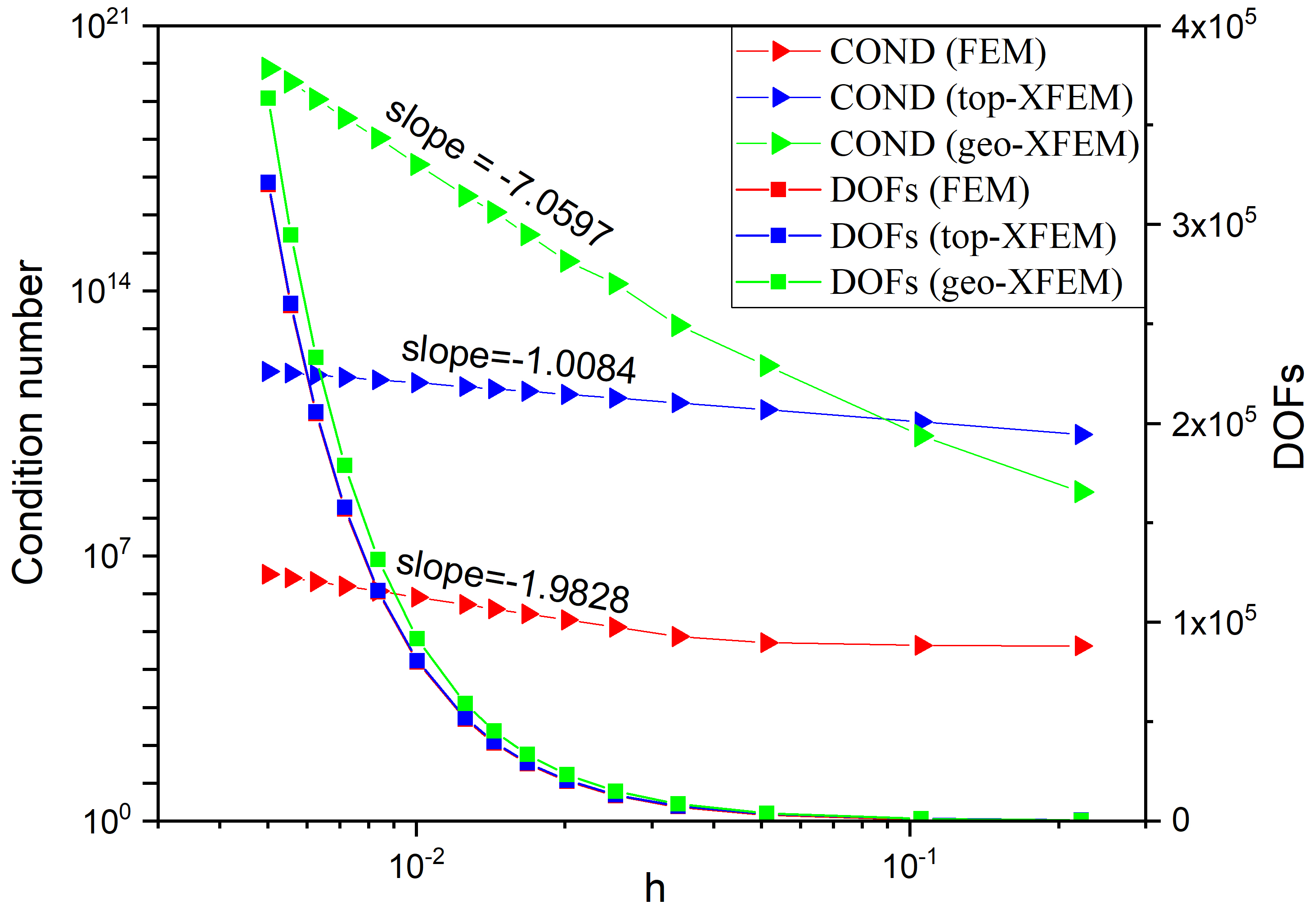}}
	\caption{Number of DOFs and condition number of the stiffness matrix for FEM and XFEM \label{fig6}}
\end{figure}

The relationship between the number of DOFs and the condition number (COND) of the stiffness matrix is plotted in Figure \ref{fig6}, demonstrating that the top-XFEM and geo-XFEM introduce some extra enriched DOFs, and the condition number is much larger than that of FEM. The number of DOFs for geo-XFEM is generally larger than that for top-XFEM because the fixed enrichment radius $r_{tip}$ enriches more layers of elements when the mesh is deeply refined. If the mesh size is coarse, the top-XFEM may generate more enriched DOFs than geo-XFEM because the actual radius of the fixed layers of elements is larger than $r_{tip}$. From the curve of the condition number, it is observed that the condition number of geo-XFEM is smaller than that of top-XFEM when the mesh is coarse, but the condition number increases rapidly and becomes much larger when the mesh is refined. According to the curve slope, the growth of the condition number with mesh refinement is $O(h^{-1.98})$ for the standard FEM, $O(h^{-1.01})$ for the top-XFEM, and $O(h^{-7.06})$ for the geo-XFEM. The results agree well with FEM theory and the numerical results of Tian et al.\cite{ref11}.

To discuss the code accuracy and convergence rate, three mesh-dependent parameters are introduced to represent the error of the numerical results, the L-2 norm, energy-norm, and SIF error. The definitions are formulated as follows.
\begin{align}
\label{eq20} \text{L2-norm}&=\sqrt{\dfrac{\sum\limits_{i=1}^{N_s}\left(u_i-u_i^h \right) \cdot\left(u_i-u_i^h \right)  }{N_s}}, \\
\label{eq21} \text{Energy-norm}&=\sqrt{\dfrac{\sum\limits_{i=1}^{N_s}\left(\varepsilon_i-\varepsilon_i^h \right):\left(\sigma_i-\sigma_i^h \right)  }{N_s}},\\
\label{eq22} \text{SIF-error}&=\frac{\left|K_I-K_I^h \right|}{\left|K_I \right|},
\end{align}
where $u,\ \varepsilon$, and $\sigma$ are the numerical displacement, strain, and stress, respectively. The superscript $h$ denotes the numerical solutions, and $N_s$ is the number of mesh points as introduced in equation (\ref{eq10}). Here, the analytical mode I stress intensity factor (SIF) is $K_I = 1.0$.

\begin{figure}[Htb] 
	\centerline{\includegraphics[scale=0.4]{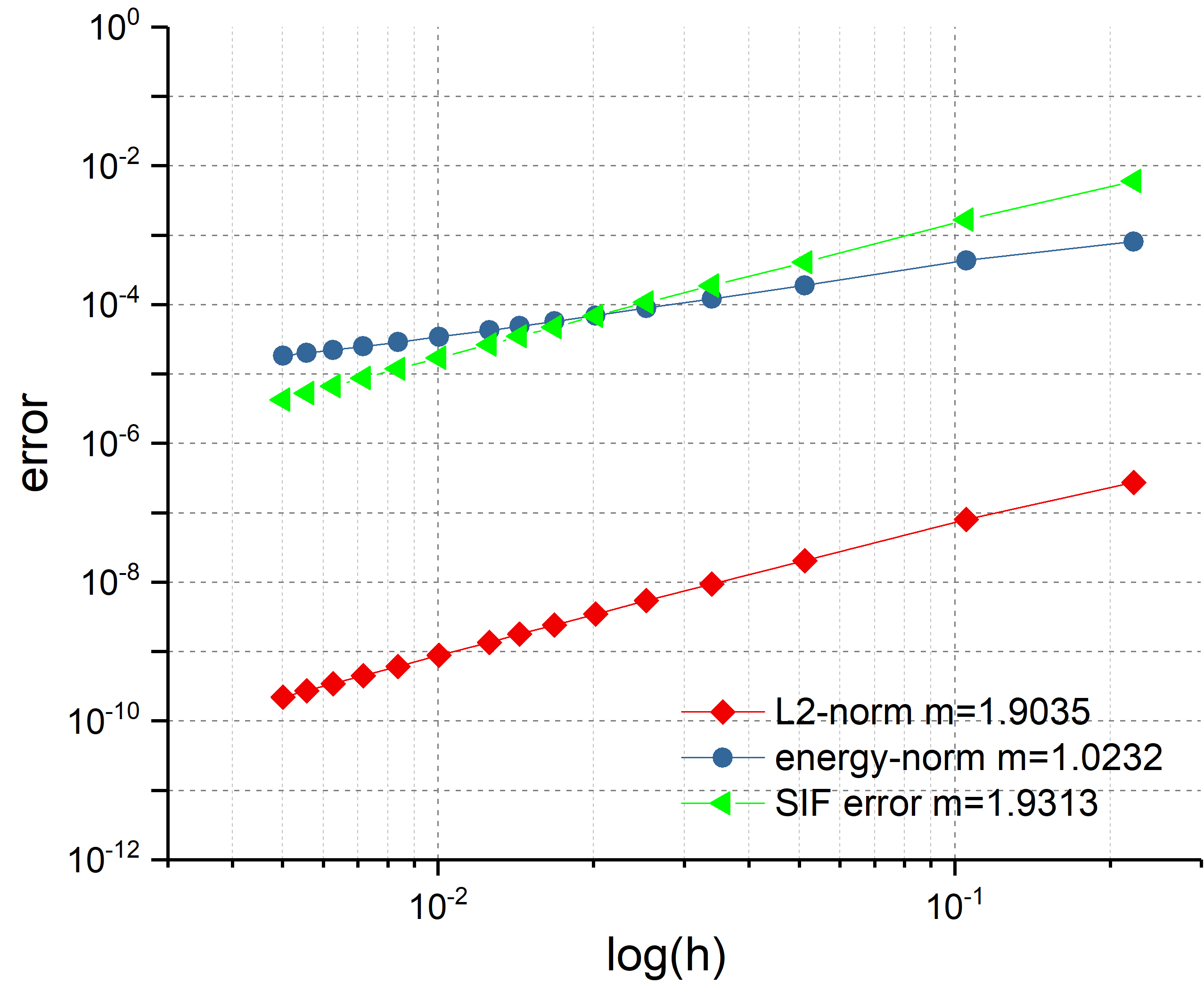}}
	\caption{Convergence order for the L2-norm, energy-norm, and SIF error with mesh refinement (m is the convergence order) \label{fig7}}
\end{figure}

Figure \ref{fig7} displays the convergence results using a log-log plot for the L2-norm, energy-norm, and SIF error. It is observed that the convergence order is 1.9035 for the L2-norm, 1.0232 for the energy-norm and 1.9313 for the SIF error, which is in good agreement with the work of Fris\cite{ref29}. 
\section{Numerical results and discussion}\label{sec4}
\subsection{Crack tip enrichment} \label{sec4.1}
\begin{figure}[Htb] 
	\centerline{\includegraphics[scale=0.10]{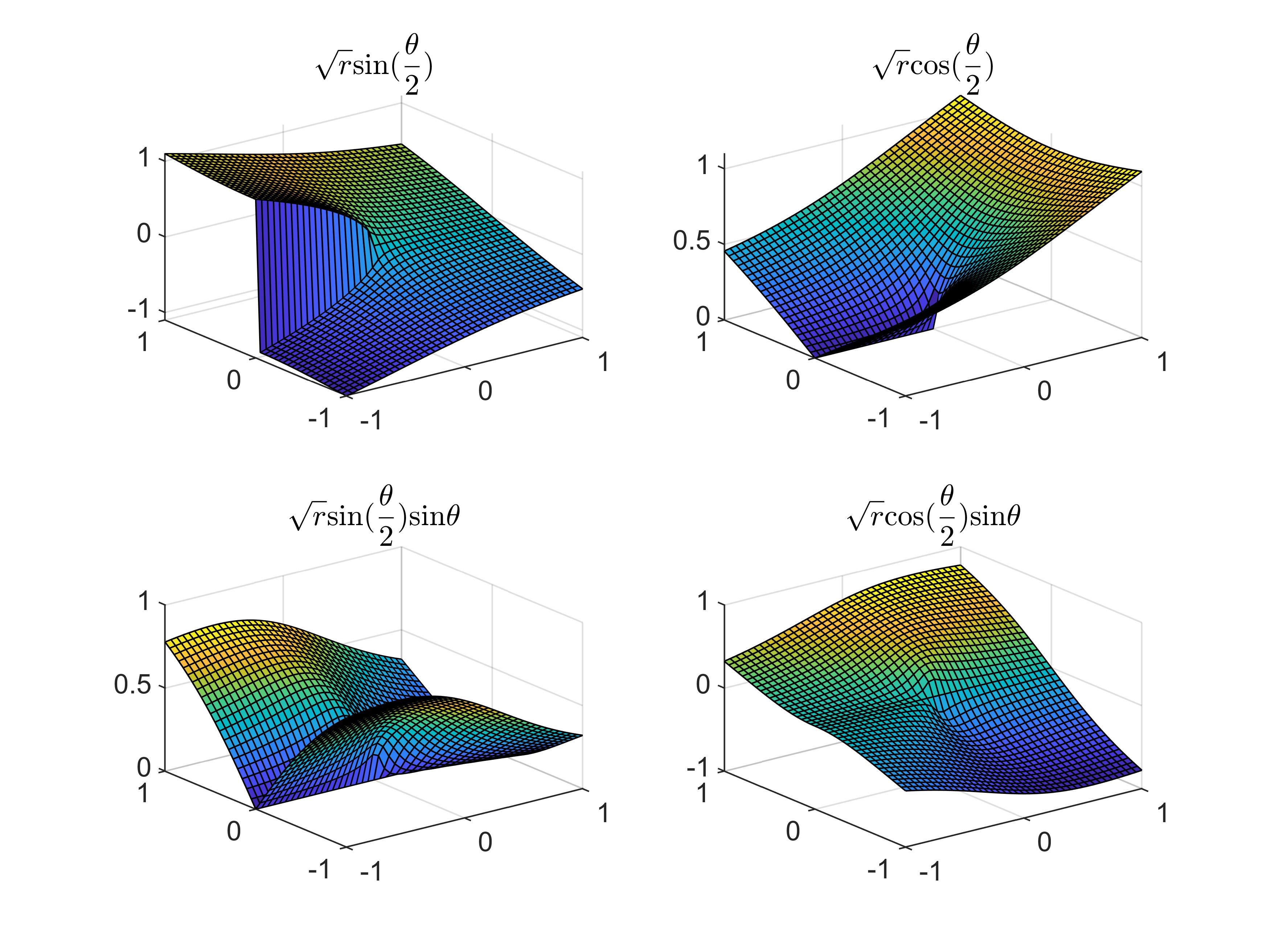}}
	\caption{Diagram of the branch enrichment functions \label{fig8}}
\end{figure}

In this section, the focus is on the crack-tip area where the difficulty arises. There are two important roles for the branch enrichment functions (\ref{eq9}): localization of where the crack is and providing the crack tip area with more information on the analytical solutions\cite{ref15}. The graphical representation of these functions is provided in Figure \ref{fig8}. It is observed that only the first term $\sqrt{r}\sin(\theta/2)$ includes a strong discontinuity, which localizes the position of the crack inside the element. The other three terms, which employ more physics in the solution, are continuous. As discussed in Section \ref{sec3}, the condition number of the stiffness matrix from the XFEM is much larger than that of the standard FEM, which presents significant challenges in problem solving. In applications, only the first few terms \cite{ref37} are adopted for simplicity, to decrease the number of DOFs and the condition number of the discrete system. The effect of adopting different terms of enrichment functions on the solver and the accuracy of the numerical results is also investigated. The crack model is the same as that described in Section \ref{sec3}.

\begin{figure}[htb] 
	\centerline{\includegraphics[scale=0.4]{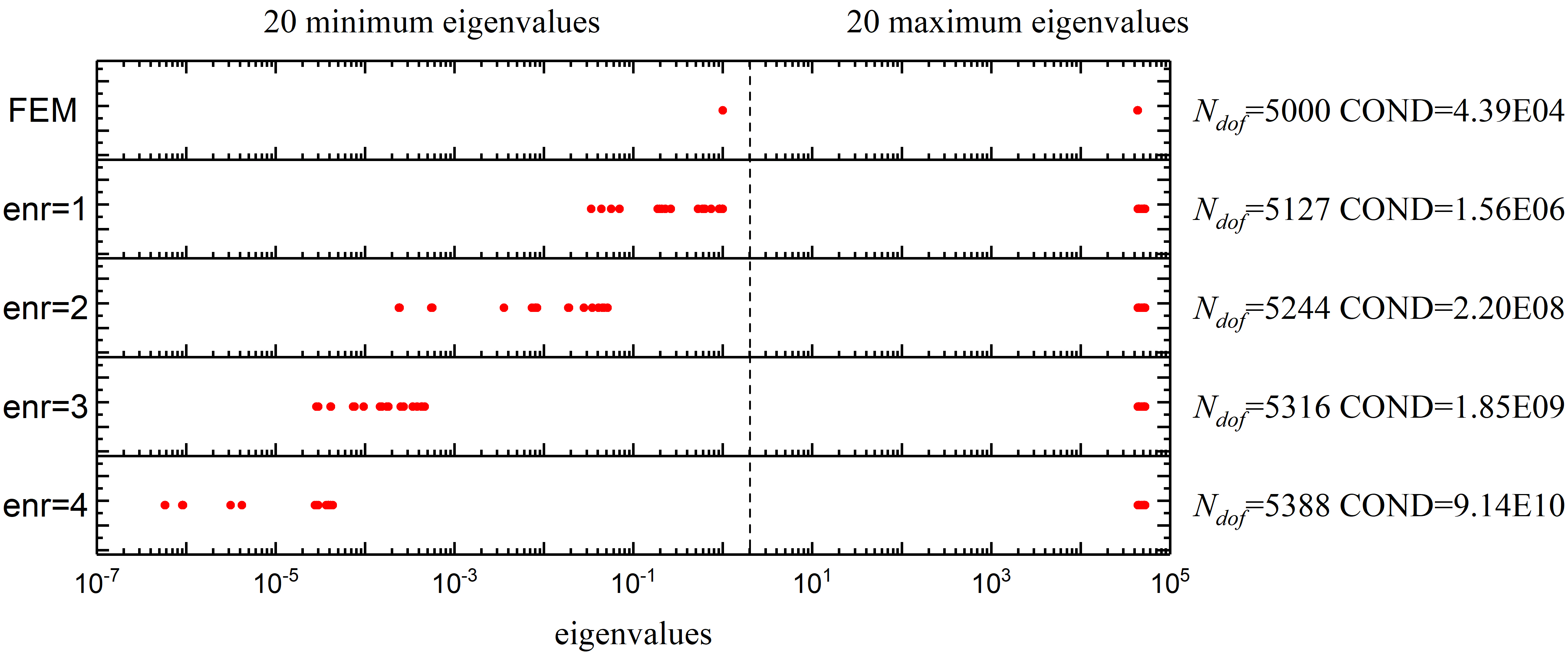}}
	\caption{Distribution of eigenvalues for different XFEMs ($N_{dof}$ is the number of DOFs. COND represents the condition number of the linear system) \label{fig9}}
\end{figure}

In Figure \ref{fig9}, the distribution of eigenvalues for different XFEMs is displayed. The number of DOFs ($N_{dof}$) and the condition number of the linear system are also compared. Here, “enr=1” indicates that the crack tip elements are enriched only by the first term of equation (\ref{eq9}); “enr=2” indicates that the first two terms are used for enrichment, and so on. It is observed that the maximum eigenvalues remain almost constant, while the minimum eigenvalues decrease quickly with increasing terms (\ref{eq9}). The condition number of a linear system is closely related to the distribution of the eigenvalues. A small minimum eigenvalue results in a large condition number. This observation suggests that the main problem arises from the branch enrichment in the crack tip area.

\begin{figure}[htb]
	\centering  
	\subfloat{ \label{fig10a}
		\begin{minipage}[t]{0.5\textwidth}
			\centering       
			\includegraphics[scale=0.36]{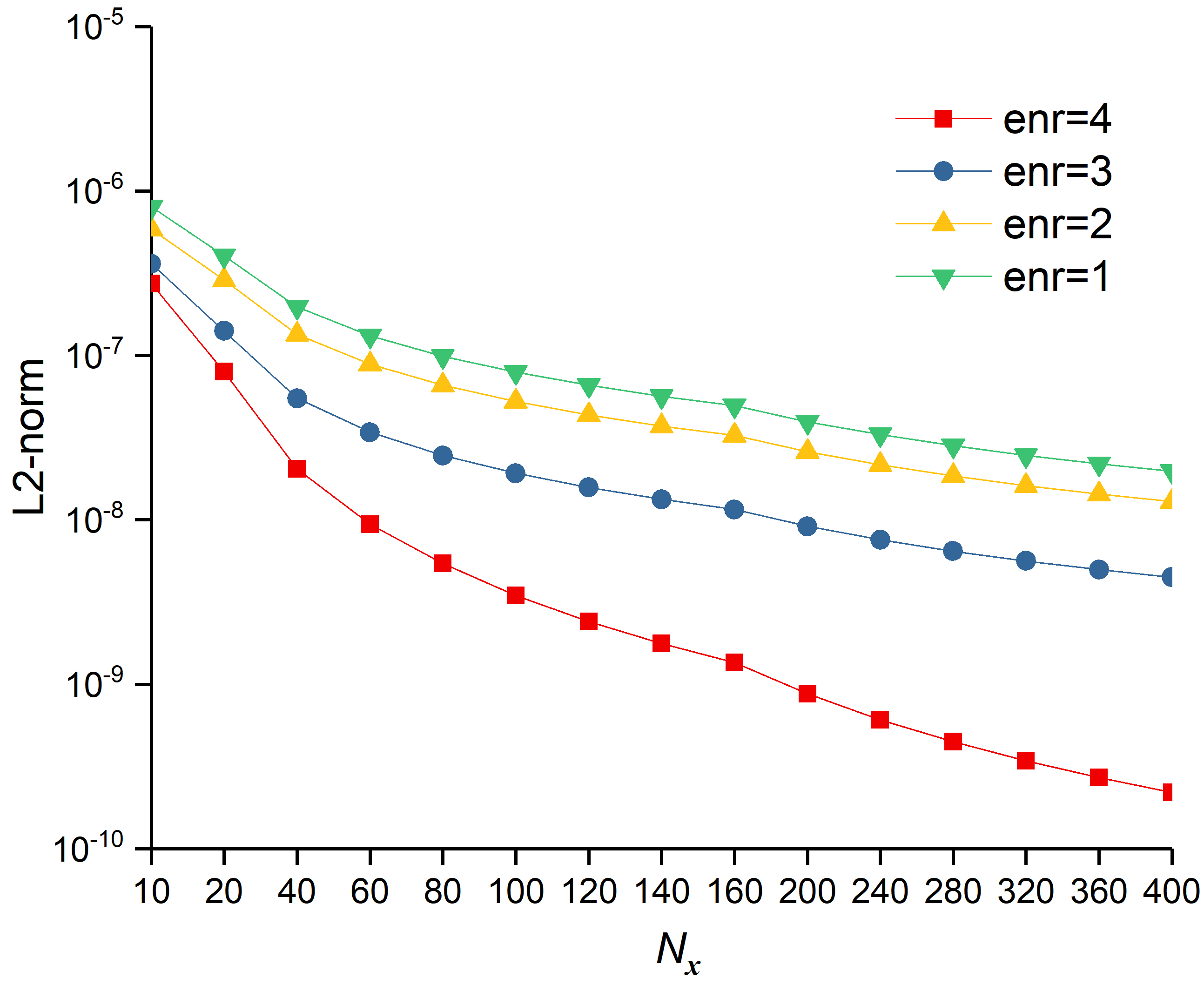}
		\end{minipage}%
	}
	\subfloat{ \label{fig10b}
		\begin{minipage}[t]{0.5\textwidth}
			\centering    
			\includegraphics[scale=0.36]{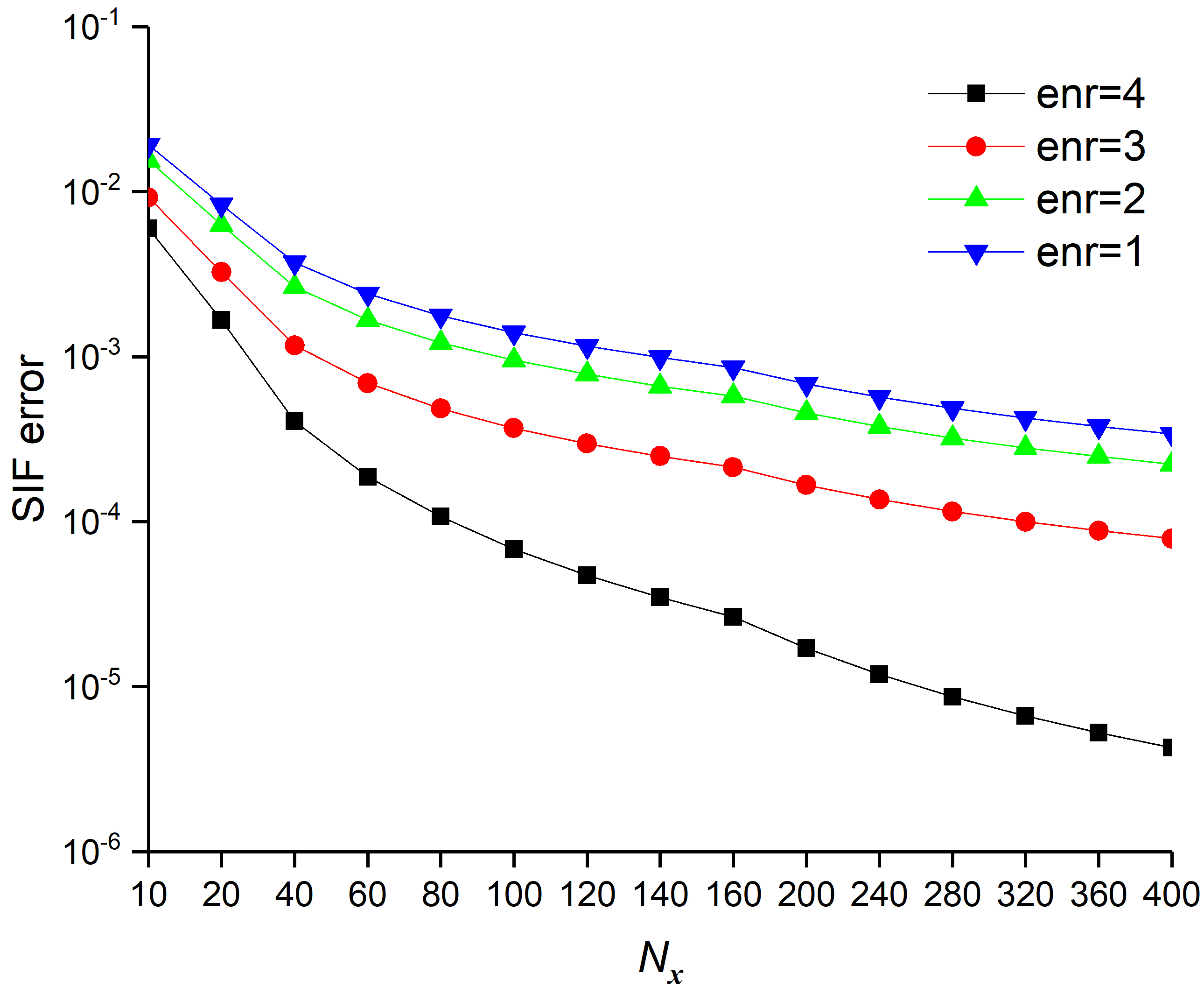}   
		\end{minipage}
	}
	\caption{Error distribution for different XFEMs with mesh refinement ($N_x$ is the number of nodes in $x$ direction)} 
	\label{fig10}  
\end{figure}

In Figure \ref{fig10}, the L2-norm and SIF errors for the four XFEMs are explored with respect to mesh refinement, whereby with the same mesh size, the full enrichment functions (enr=4) provided the most accurate results. The more the number of enrichment terms used, the better the results obtained. That is, with more enrichment terms, the basis functions in (\ref{eq9}) include more information about the analytical solution around the crack tip, and the numerical results are more credible. Table \ref{tab1} shows the convergence order for the L2-norm, energy-norm, and SIF errors. The results demonstrate that more enrichment terms yield a better convergence order. The XFEM with four enrichment functions can achieve the optimal convergence rate like the standard FEM.

\renewcommand{\arraystretch}{1.2}
\begin{table}[h]
	\centering
	\caption{Convergence order for different XFEMs \label{tab1}}%
	\begin{tabular}{|m{2cm}<{\centering}|m{1.2cm}<{\centering}|m{1.2cm}<{\centering}|m{1.2cm}<{\centering}|m{1.2cm}<{\centering}|}
		\hline
		& enr=4  & enr=3  & enr=2  & enr=1  \\ \hline
		L2-norm     & 1.9035 & 1.1434 & 1.0098 & 0.9812 \\ \hline
		energy-norm & 1.0232 & 0.3973 & 0.3993 & 0.3951 \\ \hline
		SIF error   & 1.9313 & 1.2402 & 1.1081 & 1.0578 \\ \hline
	\end{tabular}
\end{table}
\renewcommand{\arraystretch}{1}

The above discussion indicates that the full enrichment function (enr=4), apart from increasing the condition number of the linear system, also increases the accuracy of the results significantly, and provides the best convergence order. For a linear system that uses part of the branch enrichment functions for the crack problem, conventional iterative solvers can provide good performance. However, the numerical tests demonstrate that these solvers do not converge well for enr=4, unless a good preconditioner is applied to decrease the condition number. 

Note that the stiffness matrix resulting from the XFEM discretization of linear elastic problem is symmetric. After applying the boundary conditions and eliminating the linear dependency, the coefficient matrix for the linear system is symmetric positive definite (SPD) \cite{ref41}. For this type of system, the conjugate gradient (CG) algorithm is a good choice for SPD matrices \cite{ref38}. For each subdomain problem, the submatrix is also symmetric, and the complete Cholesky factorization (CC) or incomplete Cholesky factorization (ICC) is adopted as the subdomain solver. Therefore, all cases in this study are solved by CG with an ASM preconditioner for the enr=4 problem, and the subsolver for each subdomain is CC or ICC with $l$ levels of fill-ins, abbreviated as ICC($l$). 

\subsection{Comparison of subdomain strategies} \label{sec4.2}
In this section, the performances of the different subdomain strategies introduced in Section \ref{sec2.2} are compared. The branch crack model shown in Figure \ref{fig13a} is employed for the discussion. Young’s modulus $E=10^4$ MPa, and Poisson's ratio $\nu=0.30$. Initially, the condition number of the stiffness matrix is studied before and after preconditioning with a small mesh, as it is difficult to compute the condition number for large size matrices. The mesh size of 26$\times$26 is partitioned into 5$\times$5 subdomains. The overlapping size is 2. There are 1352 regular DOFs, 1088 enriched DOFs, and the assembled stiffness matrix size is 2440$\times$2440. According to the results in Table \ref{tab2}, it is observed that the condition number of the stiffness matrix is very large without the preconditioner. For the preconditioner with strategy S0, improvements do not impact the magnitude of the condition number. Comparing the results of S1 and S2, the minimum singular value and condition number with strategy S2 are noticeably smaller than those of S1. As the magnitude of the condition number decreases from $10^{11}$ to $10^5$, the domain decomposition method S2 significantly decreases the condition number. 

\renewcommand{\arraystretch}{1.2}
\begin{table}[h]
	\centering
	\caption{The singular values and condition number of $\textit{\textbf{K}}$ and $\textit{\textbf{M}}_\text{ASM}^{-1}\textit{\textbf{K}}$ for different subdomain strategies ($maxs$ is the maximum singular value; $mins$ is the minimum singular value; and $cond$ is the L2-norm condition number such that $cond = maxs/mins$) \label{tab2}}%
	\begin{tabular}{|p{1.2cm}<{\centering}|p{1.6cm}<{\centering}|p{1.6cm}<{\centering}|p{1.6cm}<{\centering}|p{1.6cm}<{\centering}|}
		\hline
		\multirow{2}{*}{\makecell[c]{}} &\multirow{2}{*}{\makecell[c]{\textit{\textbf{K}}}} & \multicolumn{3}{c|}{$\textit{\textbf{M}}_\text{ASM}^{-1}\textit{\textbf{K}}$}  \\  \cline{3-5}
		& & S0  & S1 & S2             \\ 	\hline
		$maxs$ & 8.10e+04   & 3.34e+02    	& 2.72e+03  	& 8.00e+02  	\\ 	\hline
		$mins$ & 9.58e-08  	& 1.53e-09   	& 5.68e-07 		& 4.47e-03 	\\	\hline
		$cond$ & 8.46e+11  	& 2.19e+11     	& 4.80e+09  	& 1.79e+05  	\\ 	\hline
	\end{tabular}
\end{table}
\renewcommand{\arraystretch}{1}

To compare the performance of the three subdomain strategies in parallel computing, two different meshes with mesh sizes of $1000 \times 1000$ and $5000 \times 5000$ are adopted for this discussion. The corresponding number of nodes is 1 million and 25 million, respectively. The number of processors used to solve the two problems is 24 and 1024, respectively. The overlapping size is 4 for S0, and 2 for S1 and S2. The subdomain solver of each processor for S0 and S1 is ICC(9). For S2, the subdomain solver is the complete Cholesky factorization (CC) for the tip subdomains and ICC(9) for regular subdomains. As displayed in Table \ref{tab3}, for the linear elastic crack problem, strategy S0 with the preconditioned CG method converges very slowly. When the mesh is refined to 25 million nodes, more than 4500 iterations are required for convergence. The geometrical additive Schwarz preconditioner with strategies S1 and S2 can reduce the number of iterations significantly. Compared with S1, S2 requires fewer iterations and elapsed time as it handles the crack tip subdomain separately, where the difficulties originate. Therefore, the domain decomposition preconditioner from S2 has better performance than the algebraic domain decomposition method S0 and the traditional geometrical domain decomposition method S1.

\renewcommand{\arraystretch}{1.2}
\begin{table}[h]
	\centering
	\caption{The number of iterations for different subdomain strategies (ITER is the number of iterations; $T_{sol}$ is the elapsed time of the solver) \label{tab3}}%
	\begin{tabular}{|p{1.8cm}<{\centering}|p{1.8cm}<{\centering}|p{1.8cm}<{\centering}|p{1.8cm}<{\centering}|p{1.8cm}<{\centering}|}
	\hline
	\multirow{2}{*}{\makecell[c]{subdomain \\ strategy}} & \multicolumn{2}{c|}{1 million nodes (24 CPUs)} & \multicolumn{2}{c|}{25 million nodes (1024 CPUs)}  \\  \cline{2-5}
	& ITER & $T_{sol}$ (s)              & ITER & $T_{sol}$ (s)             \\ 	\hline
	S0 & 2083 & 87.01                   & 4631 & 137.97                    \\ 	\hline
	S1 & 628  & 16.12                   & 2983 & 50.59                     \\ 	\hline
	S2 & 263  & 12.39                   & 1249 & 39.85                     \\	\hline
\end{tabular}
\end{table}
\renewcommand{\arraystretch}{1}

\subsection{Mesh-independent convergence} \label{sec4.3}
A desirable property of an algorithm is to maintain a similar convergence rate under different mesh resolutions \cite{ref46,ref47} referred to as mesh-independent convergence. In this section, the mesh-independent convergence of the additive Schwarz preconditioner is discussed by comparing the number of iterations and the norm of errors, including SIF-error, L2-norm, and energy-norm, as introduced in Section \ref{sec3}. The crack model remains the same as that of Section \ref{sec4.2}. The mesh is refined from 512$\times$512 to 4096$\times$4096, such that the mesh resolution changes from $h$ to $h/8$, where $h$ is the element size for the coarsest mesh. The number of processor cores for parallel computing is 192, and the overlapping size ranges from 1 to 8 to maintain the thickness of the overlap region between processors constant. Here, the overlapping size for the regular subdomain and crack tip subdomain is the same and the subsolver was the complete Cholesky factorization for both. To compute the norm of errors, the results of the finer mesh size $8192\times8192$ were adopted as reference. The $u_i, \varepsilon_i, \sigma_i, K_I$ in equations (\ref{eq20},\ref{eq21}, and \ref{eq22}) without superscript $h$ are the results of the finer mesh. 

The convergence performance is presented in Table \ref{tab4}. Clearly, the number of iterations changes slightly with mesh refinement. The SIF error, L2-norm, and energy norm become increasingly smaller. This demonstrates that the performance of the domain decomposition method in this study is independent of mesh resolution. It should be noted that the subsolver in Table \ref{tab4} is the complete Cholesky factorization instead of the incomplete Cholesky factorization, as our numerical experiments indicate that the subsolver ICC($l$) cannot achieve mesh-independent convergence. Although the convergence of complete Cholesky factorization is mesh-independent, it is too time-consuming and unsuitable for large-scale parallel computing. The subsolver for the regular subdomains in the following discussions is also ICC($l$).

\renewcommand{\arraystretch}{1.2}
\begin{table}[h]
	\centering
	\caption{Performance of convergence as the mesh is refined ($L/h$ is the number of mesh points in $x$ or $y$ direction; OLP is the overlapping size; ITER is the number of iterations )\label{tab4}}%
	\begin{tabular}{|m{1.2cm}<{\centering}|m{1.0cm}<{\centering}|m{1.2cm}<{\centering}|m{1.6cm}<{\centering}|m{1.6cm}<{\centering}|m{2.0cm}<{\centering}|}
		\hline
		$L/h$ & OLP & ITER  & SIF-error  & L2-norm  & Energy-norm  \\ \hline
		512   &1  	& 229   &3.92e-04	&8.77e-11	&1.49e-06 \\ \hline
		1024  &2	& 248   &1.18e-04	&2.46e-11	&6.48e-07 \\ \hline
		2048  &4 	& 238   &3.13e-05	&6.28e-12	&3.27e-07 \\ \hline
		4096  &8 	& 238   &7.06e-06	&1.71e-12	&1.82e-07 \\ \hline
	\end{tabular}
\end{table}
\renewcommand{\arraystretch}{1}

\subsection{The impact of submatrix reordering} \label{sec4.4}
\begin{figure}[htb] 
	\centering
		\subfloat[FbF] {\label{fig11a}
		\begin{minipage}[t]{0.3\textwidth}
			\centering       
			\includegraphics[scale=0.55]{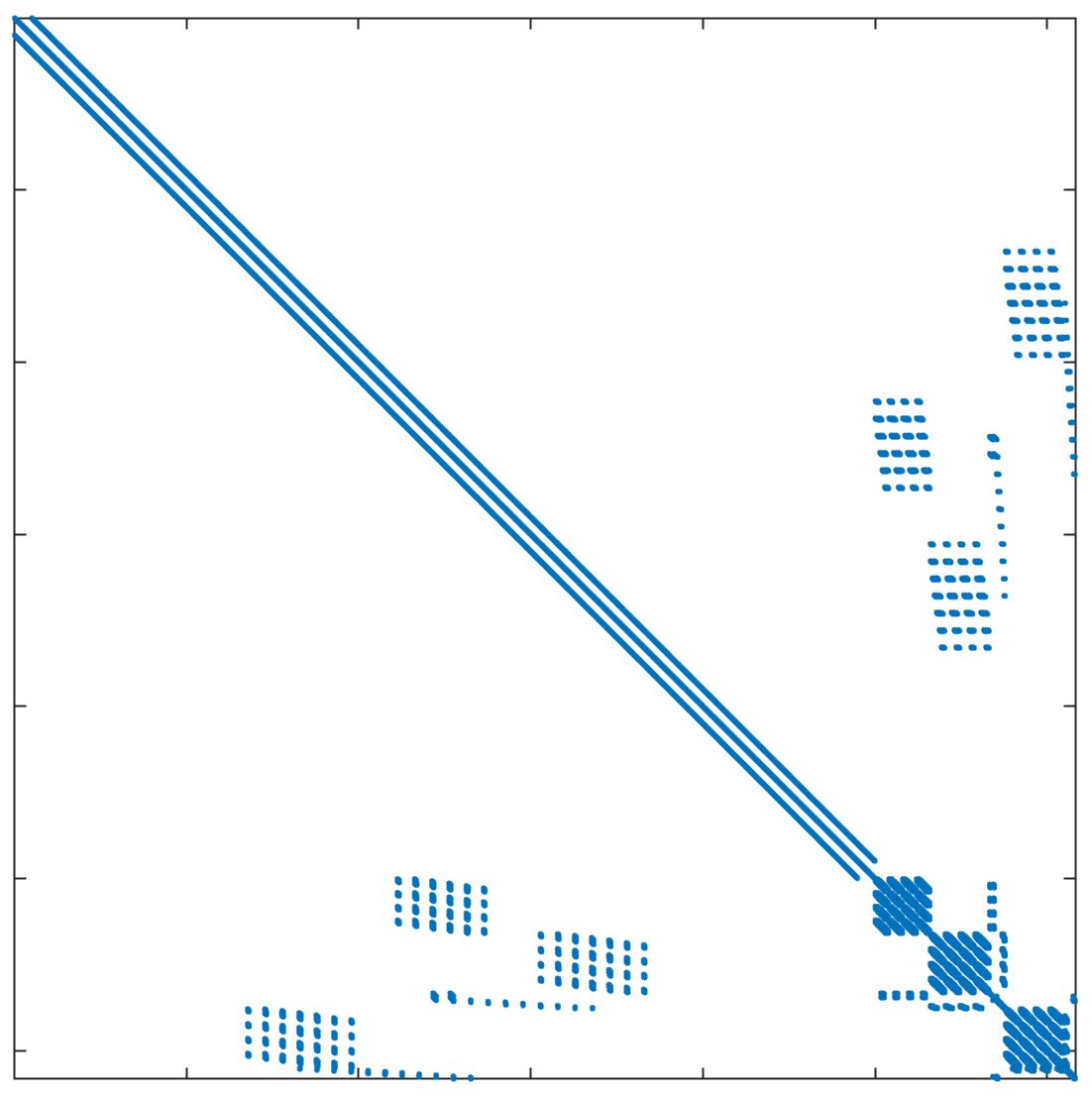}
		\end{minipage}%
	}
	\subfloat[BbB]	{\label{fig11b}
		\begin{minipage}[t]{0.3\textwidth}
			\centering    
			\includegraphics[scale=0.55]{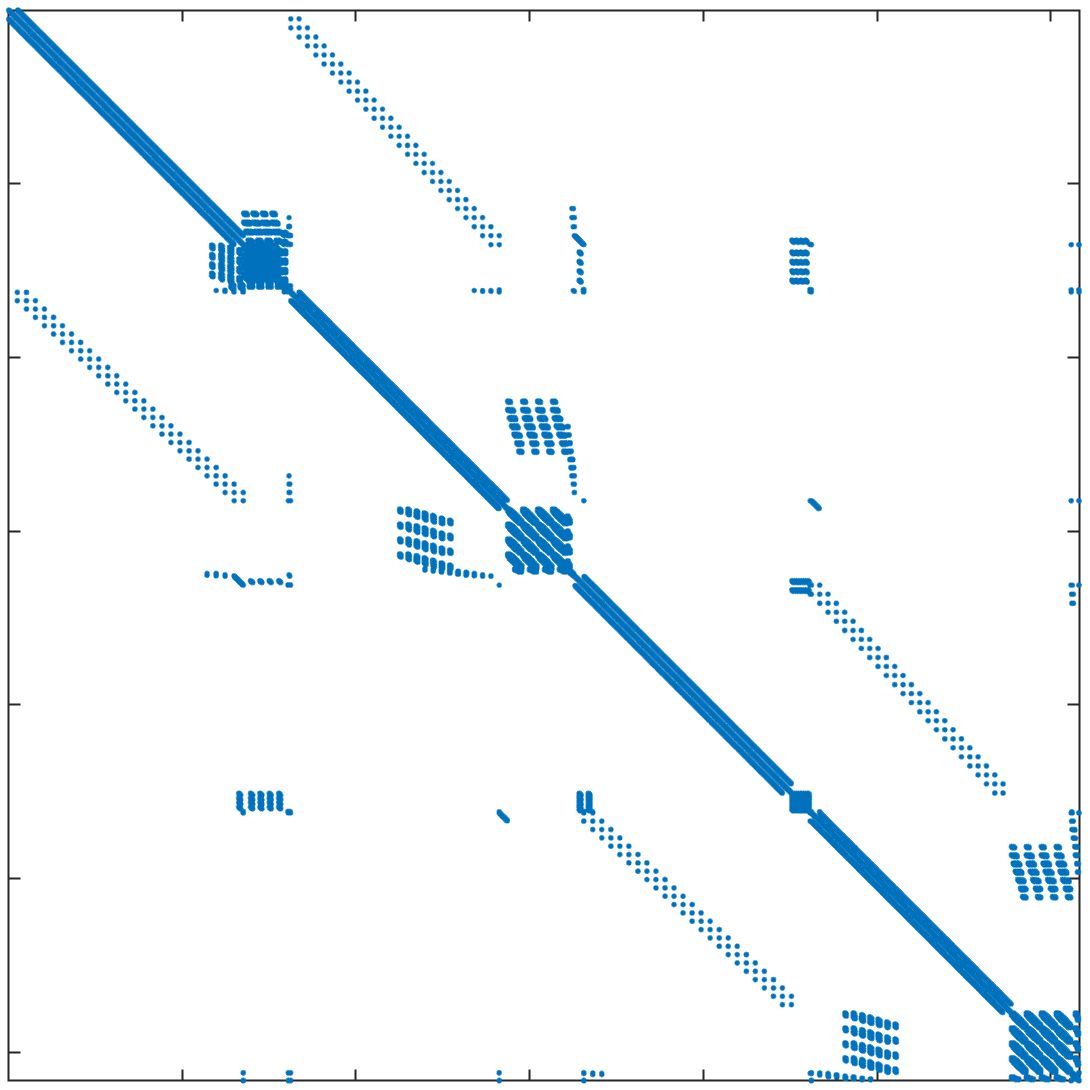}   
		\end{minipage}
	}
	\subfloat[ND]	{\label{fig11c}
		\begin{minipage}[t]{0.3\textwidth}
			\centering    
			\includegraphics[scale=0.55]{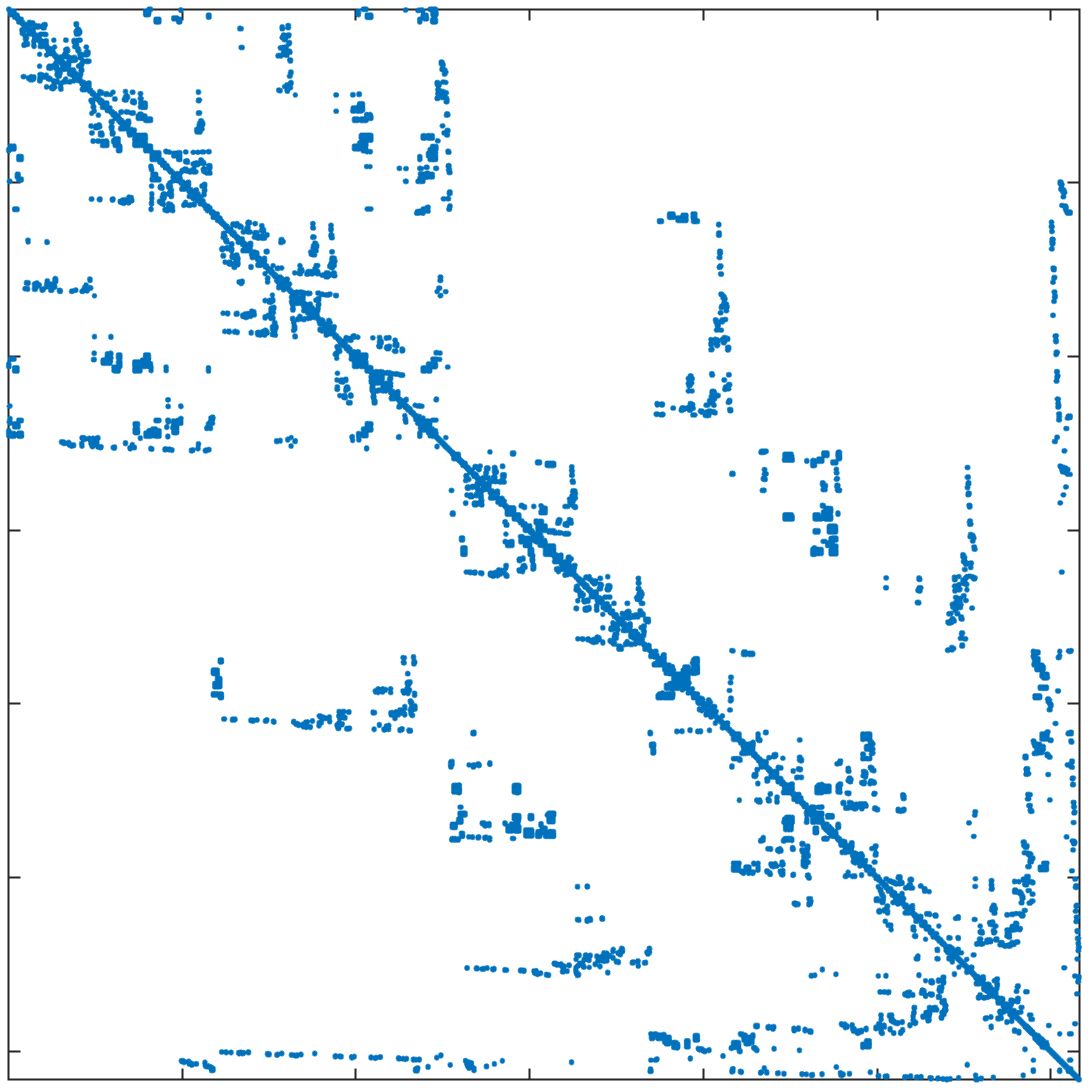}   
		\end{minipage}
	}

	\subfloat[1WD]	{\label{fig11d}
	\begin{minipage}[t]{0.3\textwidth}
		\centering    
		\includegraphics[scale=0.55]{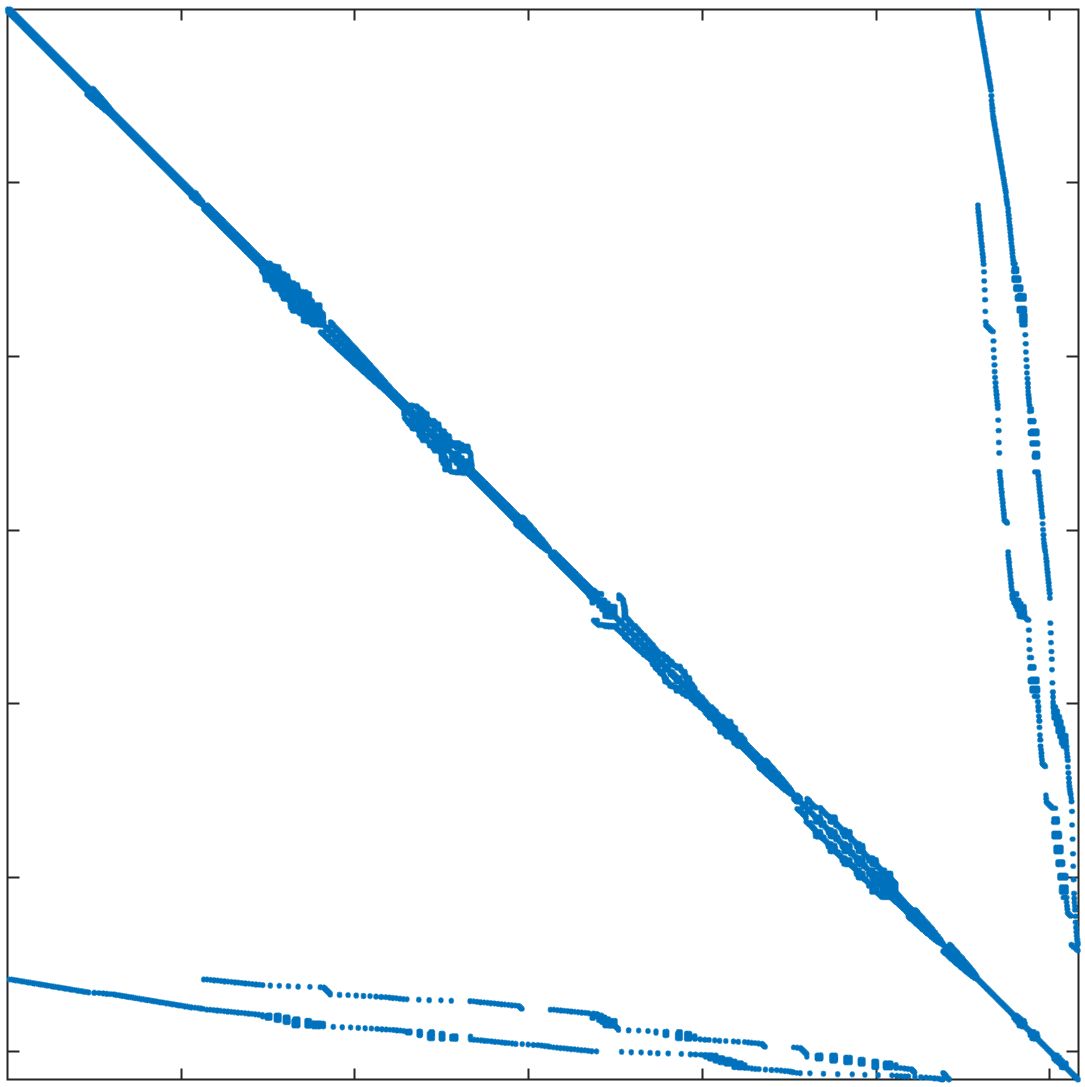}   
	\end{minipage}
}
	\subfloat[RCM]	{\label{fig11e}
	\begin{minipage}[t]{0.3\textwidth}
		\centering    
		\includegraphics[scale=0.55]{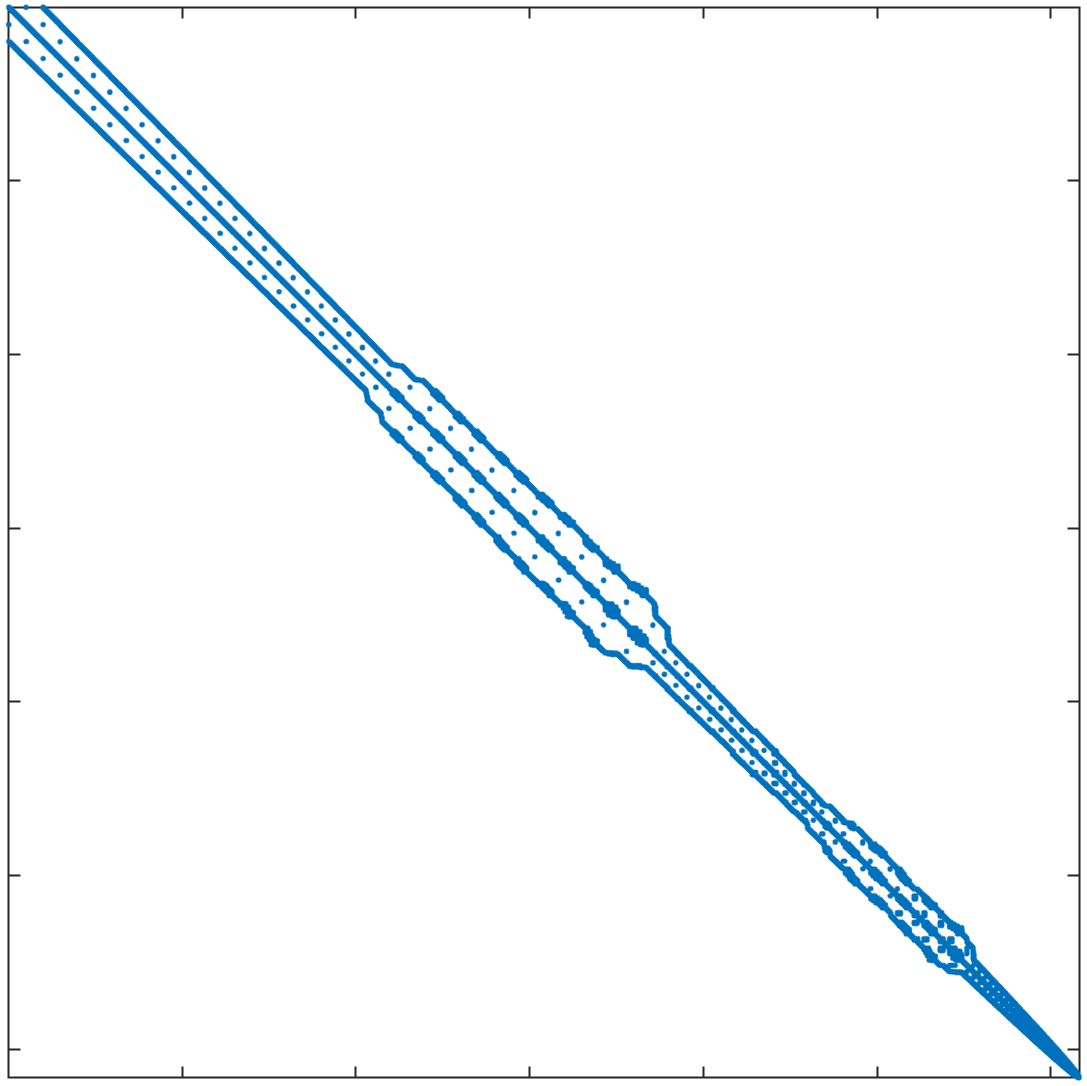}   
	\end{minipage}
}
	\subfloat[QMD]	{\label{fig11f}
	\begin{minipage}[t]{0.3\textwidth}
		\centering    
		\includegraphics[scale=0.55]{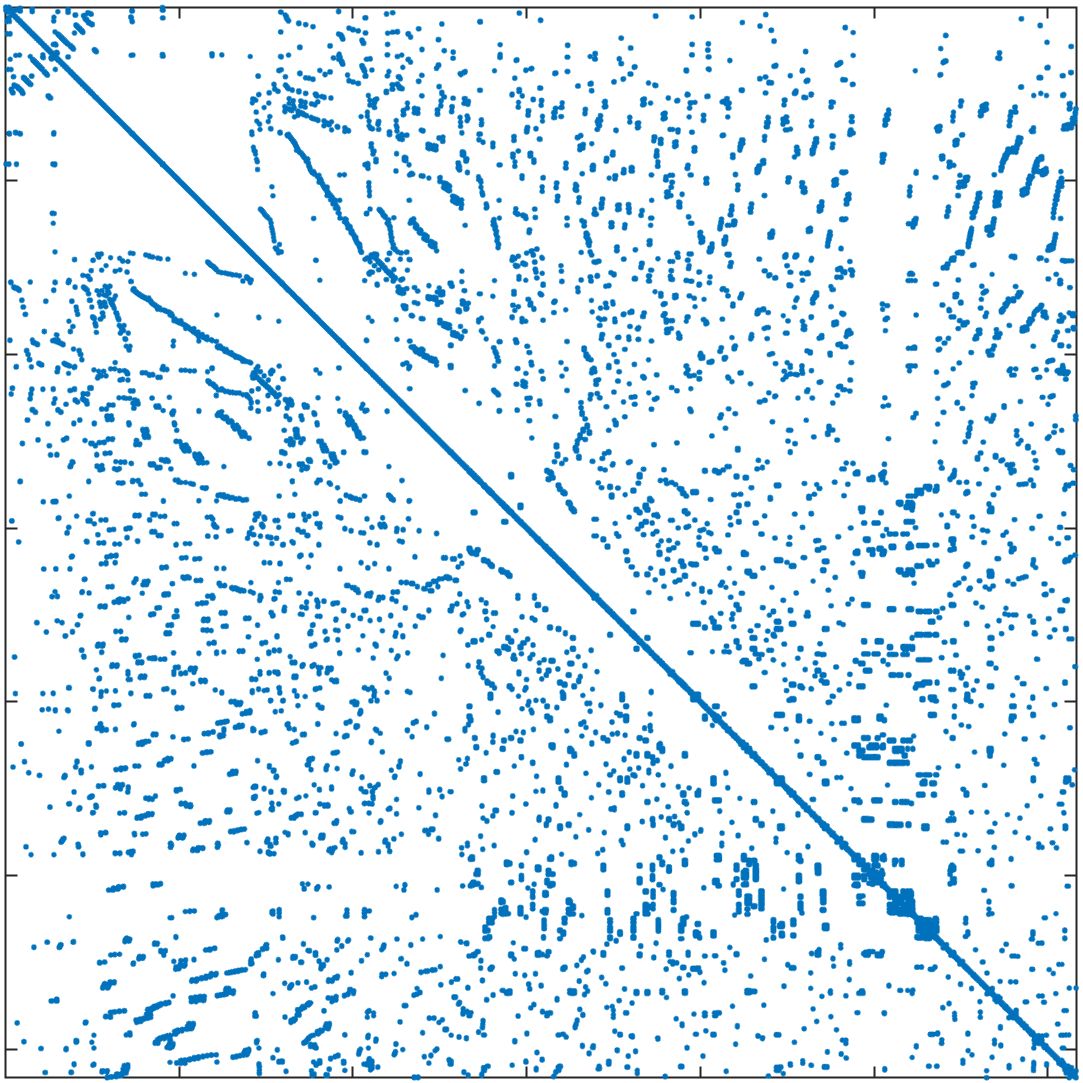}   
	\end{minipage}
}
	\caption{The nonzero structure of the stiffness matrix with different DOF ordering methods (Abbreviations: FbF, field by field; BbB, block by block; ND, nested dissection; 1WD, one-way dissection; RCM, reverse Cuthill-Mackee; QMD, quotient minimum degree. The details can be found in\cite{ref39,ref40}) \label{fig11}}
\end{figure}

In standard FEM, the DOFs are usually ordered field by field (FbF), which orders one field DOFs and then another. For the corrected XFEM, the regular DOFs are ordered first and then the crack-tip enriched DOFs and Heaviside DOFs. Figure \ref{fig11a} displays the structure of the stiffness matrix. In parallel computing, DOFs are usually ordered block by block (BbB), such that each processor orders local DOFs sequentially, and the starting index of processor $i$ is numbered after processor $i-1$ regardless of the type of DOFs. Figure \ref{fig11b} displays the structure of the stiffness matrix. It is observed that the bandwidths from the two orders are still too large for subdomain solvers such as ICC factorization \cite{ref42}. Additional reordering techniques are required to decrease the bandwidth and improve the performance of the subsolver. The matrix reordering methods include ND, 1WD, RCM, and QMD \cite{ref39,ref40}. The structure of the stiffness matrix with these four reordering methods is displayed in Figure \ref{fig11c}$\sim$\ref{fig11f}. It is observed that RCM can considerably decrease the bandwidth. The performance of the reordering techniques in parallel computing is investigated next. The crack model remains the same as in Section \ref{sec4.2} and the preconditioner is the left ASM.

In Table \ref{tab5}, the number of iterations and elapsed time of the solver are listed for the mesh with 1 million nodes and 25 million nodes. The DOFs in each processor are ordered using the BbB method by default. If the reordering method is "Natural", this means that there is no additional reordering method. The results show that compared with the natural order, ND has less elapsed time for both cases, while 1WD has no improvement and requires more iterations and elapsed time. RCM requires the least number of iterations and elapsed time among these reordering methods. QMD offers some but not tangible improvements. As RCM performs the best, it will be used as the default matrix reordering technique in the following discussions.

\renewcommand{\arraystretch}{1.2}
\begin{table}[h]
	\centering
	\caption{Impact of different submatrix reordering methods on the performance of linear solver (ITER is the number of iterations; $T_{sol}$ is the elapsed time of the solver; "Natural" means without any matrix reordering methods) \label{tab5}}%
	\begin{tabular}{|p{2.6cm}<{\centering}|p{1.6cm}<{\centering}|p{1.6cm}<{\centering}|p{1.6cm}<{\centering}|p{1.6cm}<{\centering}|}
		\hline
		\multirow{2}{*}{reordering method} & \multicolumn{2}{c|}{1 million nodes} & \multicolumn{2}{c|}{25 million nodes} \\ \cline{2-5} 
		& ITER             & $T_{sol}$ (s)              & ITER              & $T_{sol}$ (s)              \\ \hline
		Natural   &302	&15.1	&1476	&52.31		\\ \hline 
		ND        &295	&14.3	&1496	&49.02		\\ \hline
		1WD       &312	&17.25	&1510	&56.77		\\ \hline
		RCM       &263	&12.39	&1249	&39.85		\\ \hline
		QMD       &297	&15.33	&1475	&46.96		\\ \hline
	\end{tabular}
\end{table}
\renewcommand{\arraystretch}{1}

\subsection{The level of fill-ins for ICC} \label{sec4.5}
Generally, for large-scale parallel computing, complete Cholesky factorization for a symmetric sparse matrix is very time- and memory-consuming as the complexity is $O(N^3)$ for an $N\times N$ matrix \cite{ref43}. An incomplete Cholesky factorization with $l$ levels of fill-ins, abbreviated as ICC($l$), is an optional method for obtaining the subdomain preconditioner $B_i^{-1}$ in (\ref{eq15}) because the complexity is $O(N^2)$. The effect of the ICC fill-ins level on the linear solver is investigated next by comparing the number of iterations and elapsed time. The branch crack model is employed in this section; the mesh is 5000$\times$5000; and the number of unknowns is 50,015,964. The problem is computed by 1024 processors with a left ASM preconditioner; the overlapping sizes for the regular subdomains and crack tip subdomains are 2 and 6, respectively. 

\begin{figure}[htb]
	\centering  
	\subfloat[ ] {\label{fig12a}
		\begin{minipage}[t]{0.5\textwidth}
			\centering       
			\includegraphics[scale=0.35]{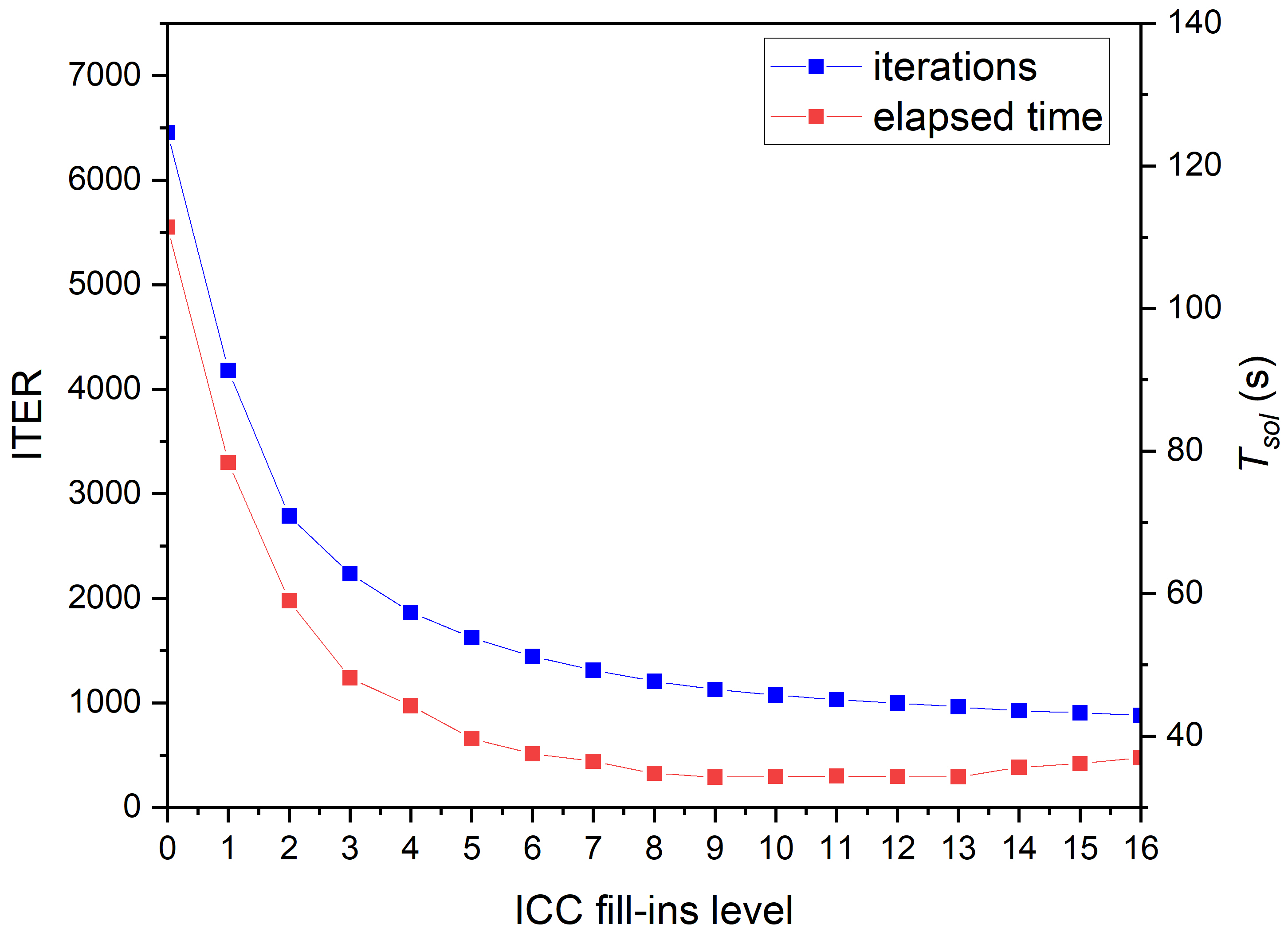}
		\end{minipage}%
	}
	\subfloat[ ]	{\label{fig12b}
		\begin{minipage}[t]{0.5\textwidth}
			\centering    
			\includegraphics[scale=0.35]{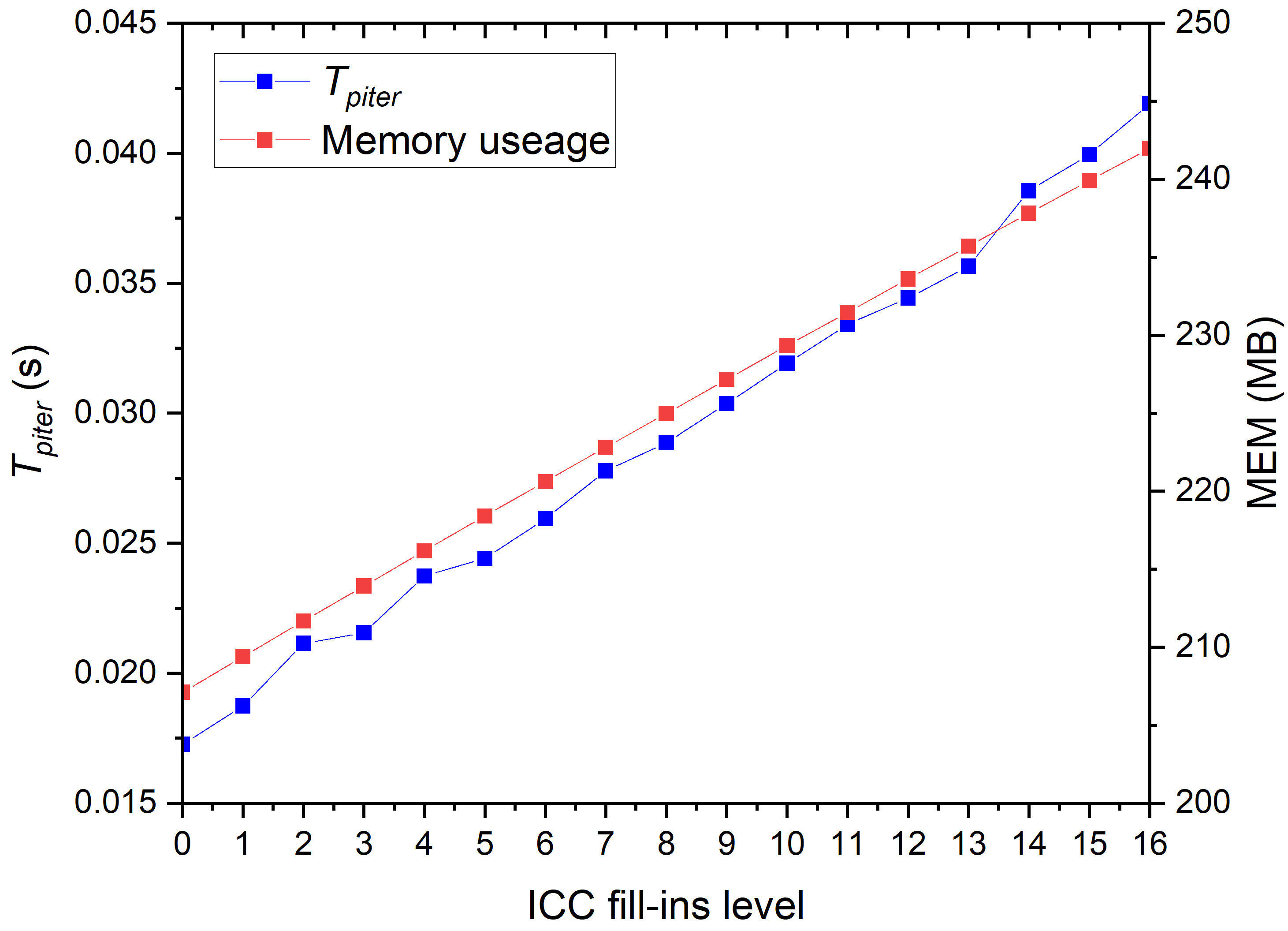}   
		\end{minipage}
	}
	\caption{The effect of ICC fill-ins level when considering regular subdomains, (a) ITER is the number of iterations; $T_{sol}$ is the elapsed time of the solver. (b) $T_{piter}$ is the elapsed time per iteration; MEM is the memory usage per processor core in megabytes.} 
	\label{fig12}  
\end{figure}

The effects of ICC fill-ins level for tip subdomains and regular subdomains are discussed separately. When considering regular subdomains, the subsolver for the crack tip subdomain is fixed as complete Cholesky factorization. When considering crack tip subdomains, the subsolver for the regular subdomain is fixed as ICC(9). The results are summarized in Table \ref{tab6} and Figure \ref{fig12}. On the left of Table \ref{tab6}, the ICC fill-ins level for the regular subdomain changes from 0 to 16 indicating that the number of iterations continues to decrease. The elapsed time of the solver decreases at first, and finally, there are some increases. Memory usage increases continuously with the increase of the ICC fill-ins level, increasing to a maximum when the subsolver is the complete Cholesky factorization. In Figure \ref{fig12a}, the number of iterations and the elapsed time drop quickly at first and then gradually when the fill-ins level is smaller than 5. When the fill-ins level is larger than 9, the elapsed time begins to increase as the elapsed time per iteration increases continuously, as shown in Figure \ref{fig12b}. This is the reason why $T_{sol}$ increases for large fill-ins level. Compared with the results of complete Cholesky factorization in the last row of Table \ref{tab6}, although the ICC($l$) requires more iterations to converge, the cost for each step is much less, which results in less total time used. On the right part of Table \ref{tab6}, the situation is different. When the ICC fill-ins level for the crack tip subdomain is less than 7, the solver does not converge. With the increase of the ICC fill-ins level, the solver converges and quickly attains the results of CC. This indicates that the crack tip subdomain solver is much more sensitive to the accuracy of matrix factorization because the crack tip submatrix is singular. The iterations, elapsed time, and memory usage are remarkably close to those of complete Cholesky factorization in the last row. As the crack tip submatrix is denser, there is no significant difference between complete Cholesky factorization and incomplete Cholesky factorization with large fill-ins level in terms of time consumption and memory usage. For these reasons, the subsolver for the tip subdomain is set as CC, and that for the regular subdomain is ICC(9), which achieves the best performance.

\renewcommand{\arraystretch}{1.2}
\begin{table}[h]
	\centering
	\caption{Effect of ICC fill-ins level for regular subdomains and tip subdomains ("-" means not converge; the overlapping size is 2) \label{tab6}}%
	\begin{tabular}{|m{1.0cm}<{\centering}|m{1.2cm}<{\centering}|m{1.2cm}<{\centering}|m{1.2cm}<{\centering}|m{1.7cm}<{\centering}|m{1.2cm}<{\centering}|p{1.2cm}<{\centering}|m{1.2cm}<{\centering}|m{1.7cm}<{\centering}|}
		\hline
		\multirow{2}{*}{$l$} & \multicolumn{4}{c|}{\begin{tabular}[c]{@{}c@{}}tip subdomain: CC\\ reg subdomain: ICC($l$)\end{tabular}} & \multicolumn{4}{c|}{\begin{tabular}[c]{@{}c@{}}tip subdomain: ICC($l$)\\ reg subdomain: ICC(9)\end{tabular}} \\ \cline{2-9} 
		& ITER                        & $T_{sol}$ (s)                        & $T_{fac}$ (s)        &MEM (MB)               & ITER                         & $T_{sol}$ (s)                         & $T_{fac}$ (s)         &MEM (MB)                  \\ \hline
		0  		&6454		&111.43    	&0.069    	&207.11 		&-	  	&-		&-	   		&-			\\ \hline
		1  		&4182		&78.39 		&0.092 		&209.40 		&-		&-		&-			&-			\\ \hline
		2  		&2789		&58.99 		&0.112     	&211.67 		&-		&-		&-	 		&-			\\ \hline
		3  		&2235		&48.18 		&0.141     	&213.93 		&-		&-		&-	 		&-			\\ \hline
		4		&1865		&44.27 		&0.167 		&216.17 		&-		&-		&-			&-			\\ \hline
		5  		&1625		&39.67 		&0.201     	&218.40 		&-		&-		&-	 		&-			\\ \hline
		6		&1447		&37.54 		&0.244 		&220.61 		&-		&-		&-			&-			\\ \hline
		7  		&1313		&36.48 		&0.295     	&222.81 		&-		&-		&-			&-			\\ \hline
		8		&1206		&34.80 		&0.348 		&225.00 		&1400	&43.03 	&0.028		&226.99		\\ \hline
		9  		&1128		&34.26 		&0.386     	&227.17 		&1381	&41.86 	&0.031		&227.02		\\ \hline
		10		&1076		&34.35 		&0.445 		&229.33 		&1235	&37.55 	&0.032		&227.04		\\ \hline
		11 		&1030		&34.42 		&0.511     	&231.47 		&1142	&35.33 	&0.033		&227.06		\\ \hline
		12		&998		&34.36 		&0.563 		&233.60 		&1139	&34.53 	&0.038		&227.08		\\ \hline
		13 		&962		&34.30 		&0.625     	&235.71 		&1136	&34.40 	&0.041		&227.08		\\ \hline
		14		&924		&35.62 		&0.689 		&237.82 		&1128	&33.92 	&0.042		&227.10		\\ \hline
		15 		&905		&36.16 		&0.765     	&239.90 		&1123	&33.87 	&0.045		&227.11		\\ \hline
		16		&883		&37.01 		&0.854 		&241.98 		&1128	&34.53 	&0.046		&227.12		\\ \hline
		CC		&739		&128.24 	&25.036     &449.57 		&1128	&34.10 	&0.054		&227.17		\\ \hline
	\end{tabular}
\begin{tablenotes}
	\item Abbreviation: $l$ is the ICC fill-ins level; ITER is the number of iterations; $T_{sol}$ is the elapsed time for solver; $T_{fac}$ is the elapsed time for submatrix factorization; MEM is the memory usage per processor core in megabytes; CC means complete Cholesky factorization.
\end{tablenotes}
\end{table}
\renewcommand{\arraystretch}{1}

\subsection{The overlapping size} \label{sec4.6}
In this section, the impact of overlapping size, which plays a critical role in the additive Schwarz method is studied. Overlap in the additive Schwarz method represents the communication between processor cores. Generally, a larger overlapping size can decrease the number of iterations because the local processor can obtain more information from neighboring processors. Meanwhile, a larger overlapping size increases the cost of the subsolver for local problems. Therefore, an appropriate overlapping size should be used to achieve the best performance. The crack model is a branch crack; the mesh size is 5000$\times$5000; and the problem is solved using 1024 processors. The subsolver for regular subdomains is ICC(9), and the crack tip subdomain is CC.

Table \ref{tab7} shows the effect of overlapping size in terms of iterations and elapsed time. Here, the overlapping size for regular subdomains and tip subdomains is discussed separately. On the left part of Table \ref{tab7}, the overlapping size for tip subdomains is fixed as 6, and that for regular subdomains varies from 1 to 7. The number of iterations keeps decreasing continuously, while the elapsed time drops at first and then begins to increase when the overlapping size is larger than 2, as the elapsed time per iteration is always increasing. It is known that each layer of overlapping elements in the geometrical domain decomposition method includes many DOFs, increasing the communication between processors. Therefore, the overlapping size for the regular subdomain can be set to OLP(2). On the right part of Table \ref{tab7}, the overlapping size for regular subdomains is fixed as 2, and that for crack tip subdomains varies from 1 to 7. With the increasing overlapping size, the iterations always decrease and the elapsed time keeps decreasing as well when the overlapping size is smaller than 6. When the overlapping size is 7, there is little increase in elapsed time. The elapsed time per iteration exhibits little change when the overlapping size increases, which can be explained by the crack tip subdomain being usually much smaller than the regular subdomain, leading to much less DOFs from crack tip overlap. The main communication during iterations mainly comes from regular subdomains. The overlapping size for crack tip subdomain has an effect on the number of iterations but not much on the elapsed time per iteration. As summarized, the overlapping size for regular subdomain is chosen as OLP(2) and for the crack tip subdomain, it is OLP(6).

\renewcommand{\arraystretch}{1.2}
\begin{table}[h]
	\centering
	\caption{Effect of overlapping size on tip subdomains and regular subdomains (the subsolver is ICC(9) for regular subdomains and CC for crack tip subdomains, the mesh size is 5000$\times$5000) \label{tab7}}%
	\begin{tabular}{|m{1.0cm}<{\centering}|m{1.2cm}<{\centering}|m{1.2cm}<{\centering}|m{1.4cm}<{\centering}|m{1.2cm}<{\centering}|m{1.2cm}<{\centering}|p{1.4cm}<{\centering}|}
		\hline
		& \multicolumn{3}{c|}{OLP(tip)=6, OLP(reg) varies} & \multicolumn{3}{c|}{OLP(reg)=2, OLP(tip) varies} \\ \hline
		OLP & ITER     & $T_{sol} (s)$       & $T_{piter} (s)$        & ITER     & $T_{sol} (s)$       & $T_{piter} (s)$        \\ \hline
		1	&1232	&35.99 	&0.029    &1304	&39.17 	&0.030   \\ \hline
		2	&1126	&33.60 	&0.030    &1250	&37.24 	&0.030   \\ \hline
		3	&1098	&34.52 	&0.031    &1185	&35.31 	&0.030   \\ \hline
		4	&1080	&34.26 	&0.032    &1150	&35.12 	&0.031   \\ \hline
		5	&1070	&34.62 	&0.032    &1138	&34.50 	&0.030   \\ \hline
		6	&1058	&34.59 	&0.033    &1128	&33.65 	&0.030   \\ \hline
		7	&1041	&36.40 	&0.035    &1130	&34.02 	&0.030   \\ \hline
	\end{tabular}
	\begin{tablenotes}
		\item Abbreviation: OLP is overlapping size; ITER is the number of iterations; $T_{sol}$ is the total elapsed time of solver; $T_{piter}$ is elapsed time per iteration.
	\end{tablenotes}
\end{table}
\renewcommand{\arraystretch}{1}

\subsection{Parallel scalability analysis} \label{sec4.7}
\begin{figure}[htb]
	\centering  
	\subfloat[] {\label{fig13a}
		\begin{minipage}[t]{0.5\textwidth}
			\centering       
			\includegraphics[scale=0.2]{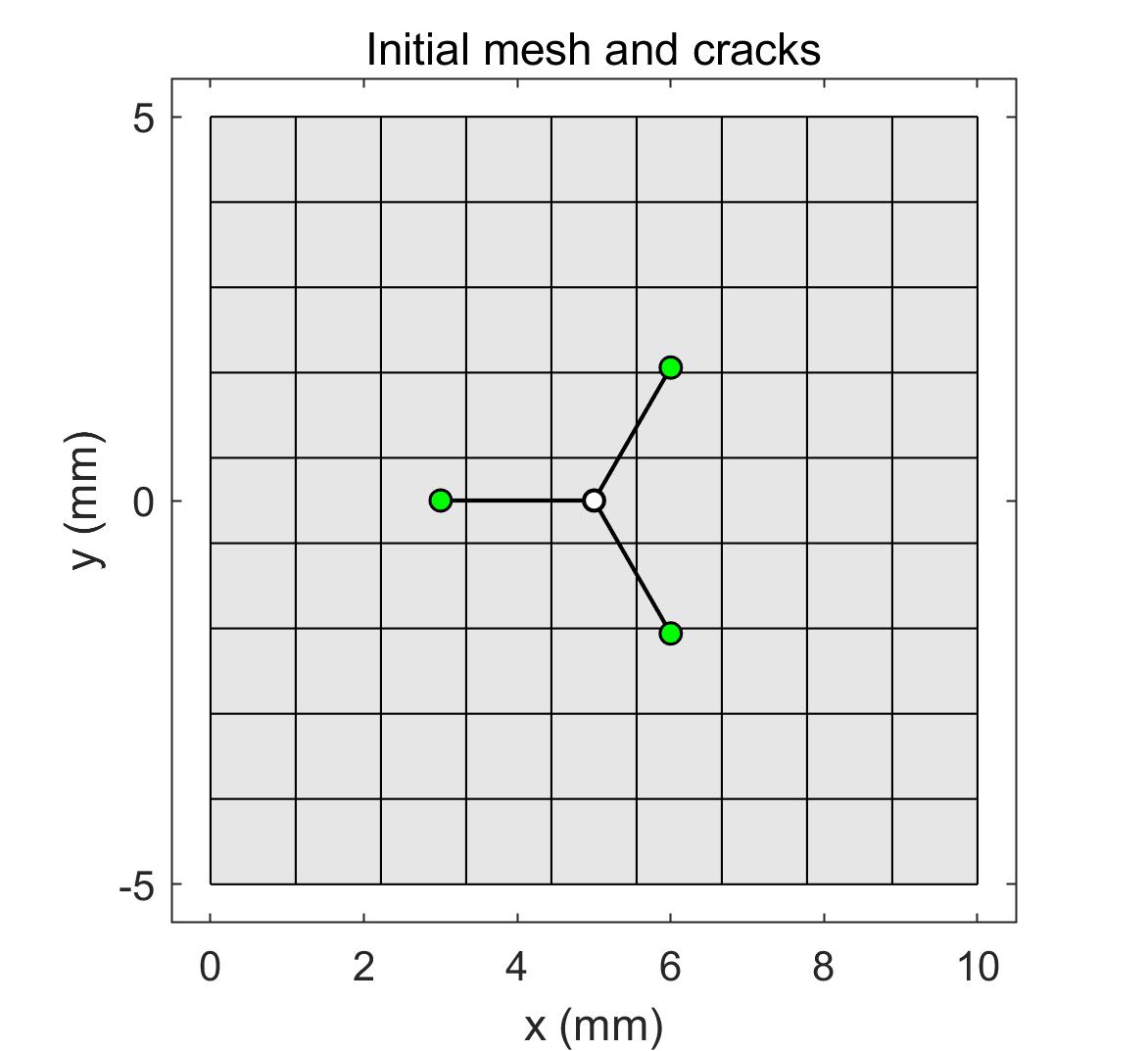}
		\end{minipage}%
	}
	\subfloat[]	{\label{fig13b}
		\begin{minipage}[t]{0.5\textwidth}
			\centering    
			\includegraphics[scale=0.2]{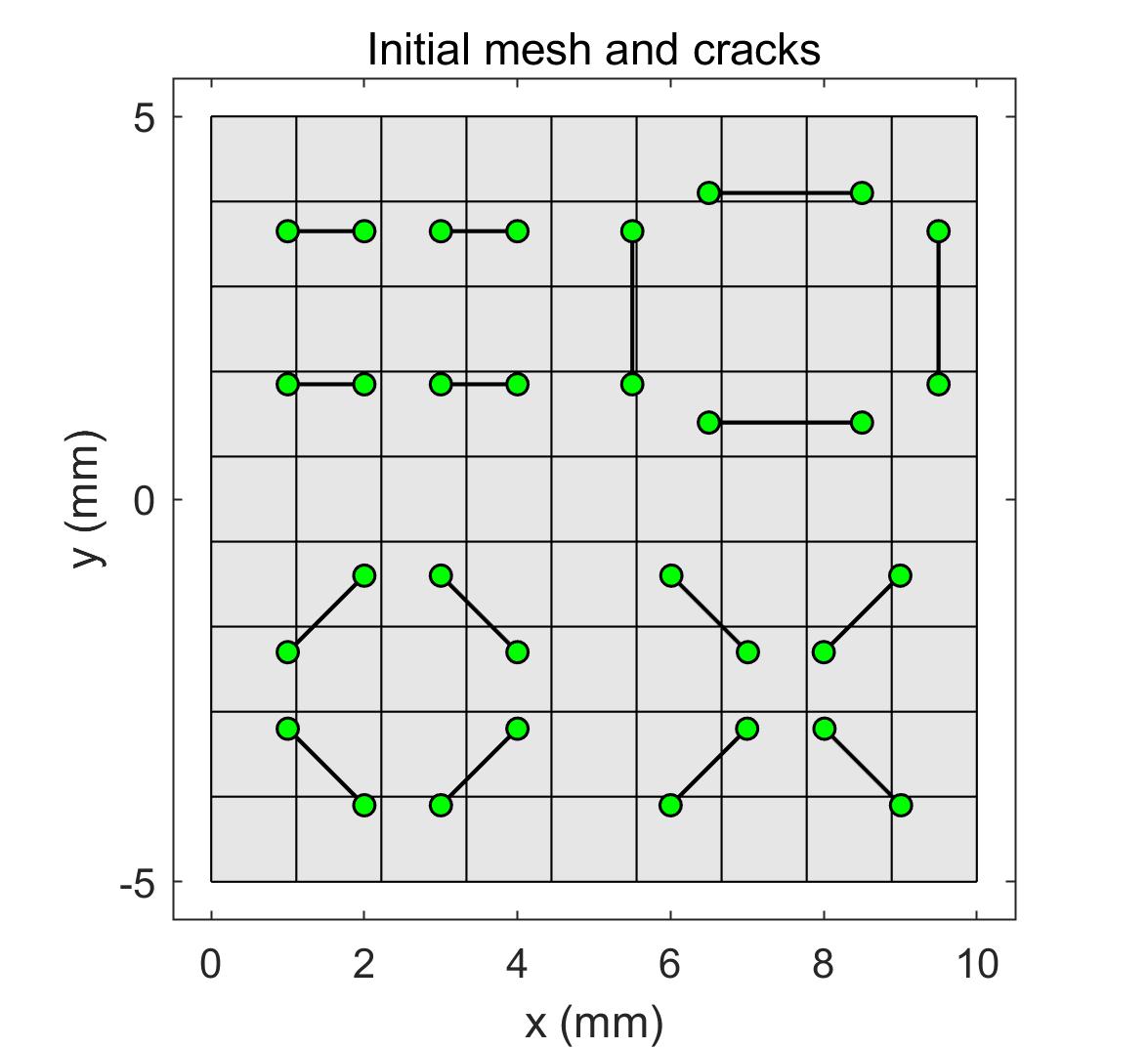}   
		\end{minipage}
	}
	\caption{Diagram for (a) branch crack model and (b) 16-cracks model (for drawing convenience, the mesh is coarsened to 10$\times$10)} 
	\label{fig13}  
\end{figure}

\begin{figure}[htb]
	\centering  
	\subfloat[ ] {\label{fig14a}
		\begin{minipage}[t]{0.5\textwidth}
			\centering       
			\includegraphics[scale=0.18]{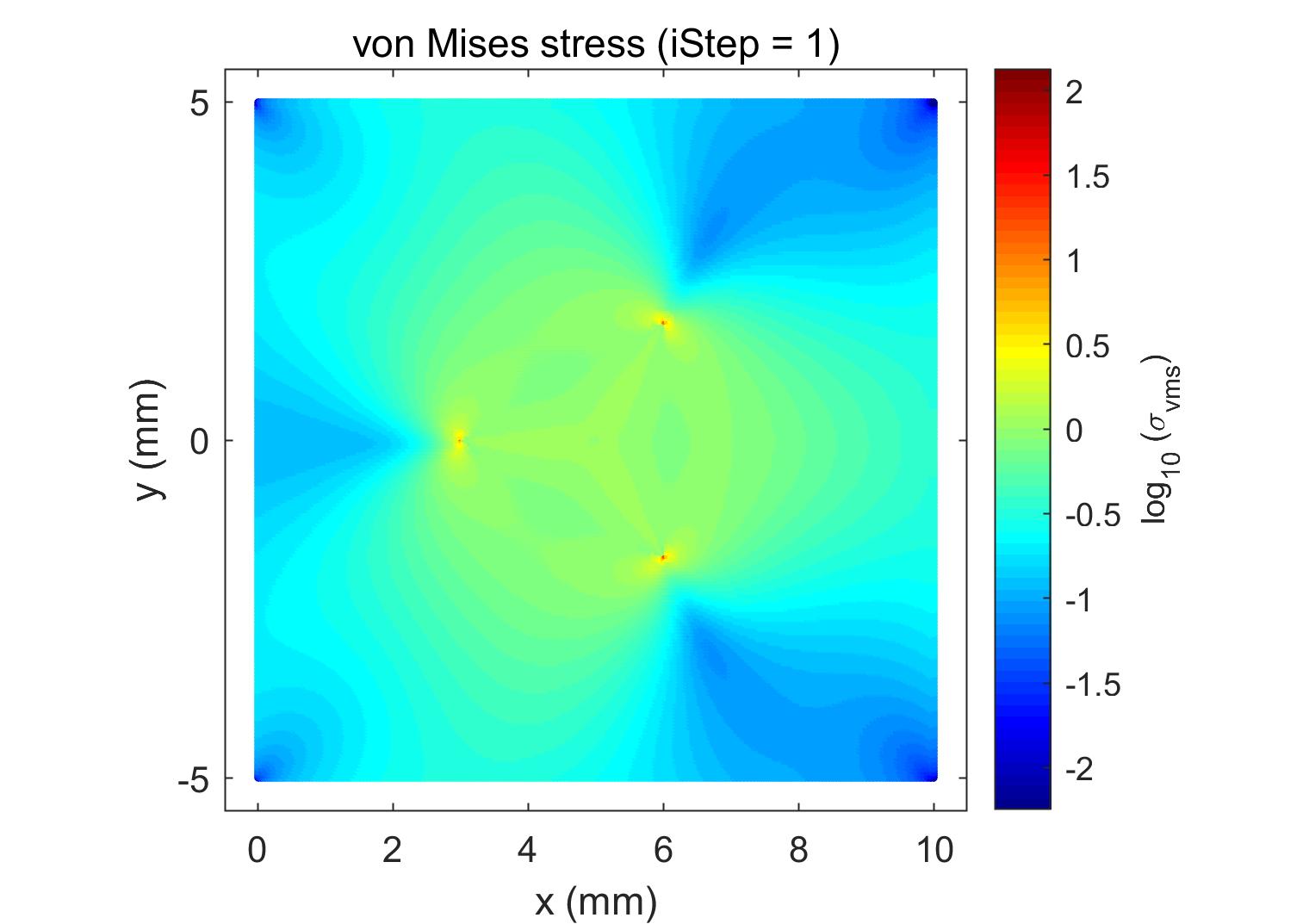}
		\end{minipage}%
	}
	\subfloat[ ]	{\label{fig14b}
		\begin{minipage}[t]{0.5\textwidth}
			\centering    
			\includegraphics[scale=0.18]{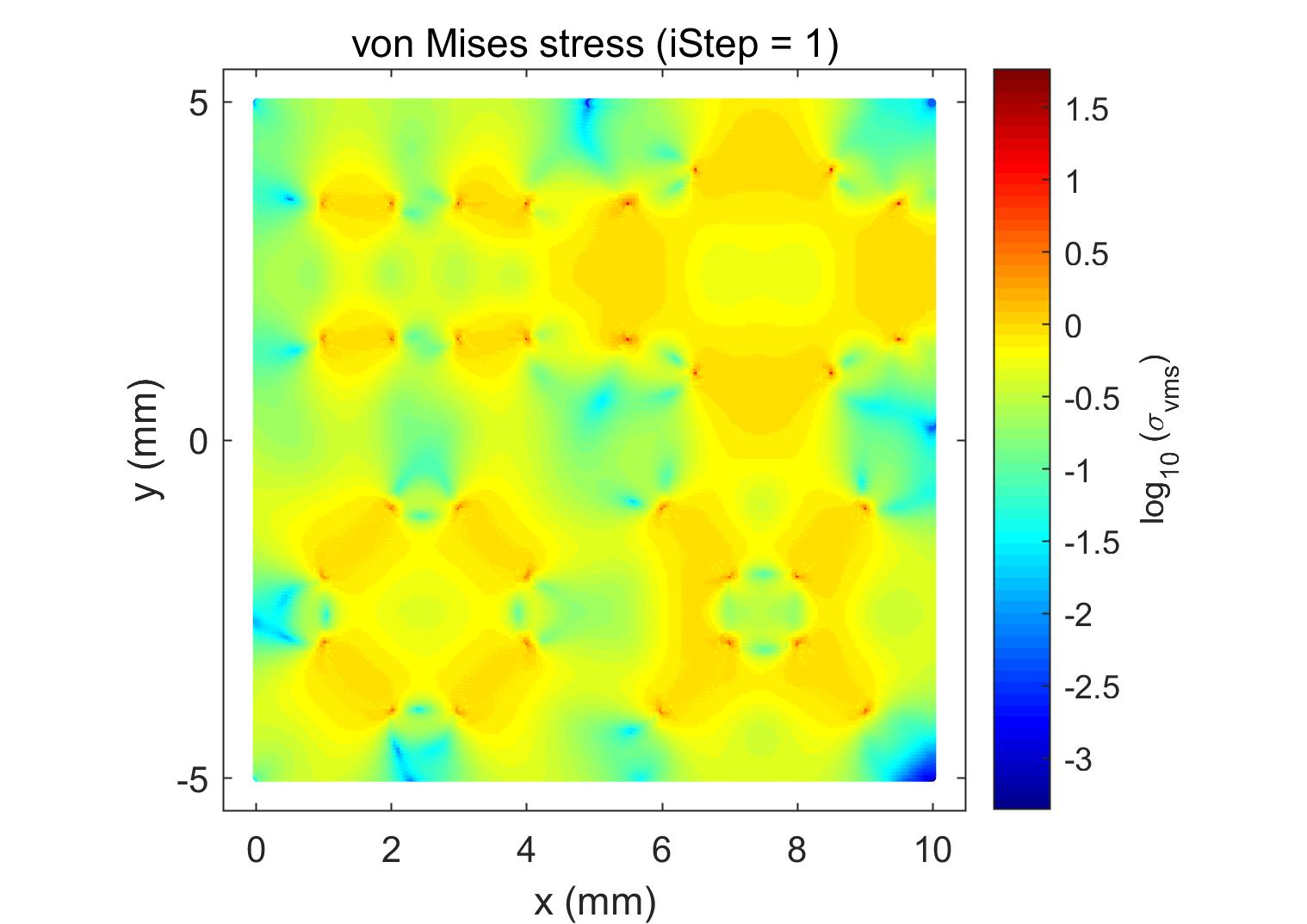}   
		\end{minipage}
	}
	\caption{The Von-Mises stress contour for (a) branch crack model and (b) 16-cracks model} 
	\label{fig14}  
\end{figure}

In this section, the parallel scalability of the proposed algorithm is discussed. As shown in Figure \ref{fig13}, a square domain of size 10 m$\times$10 m with cracks inside is considered. The first crack model is a branch crack with three crack tips. The second crack model has 16 cracks in the domain and 32 crack tips. The problem type is plain stress. Young’s modulus $E=10000$ MPa, and Poisson's ratio $\nu=0.30$. The boundaries of the domain are fixed with zero displacement. The crack surface is compressed with 1.0 MPa inner water pressure. The relative convergence tolerance was $10^{-6}$. The submatrix in each processor for all cases was reordered by RCM. The numerical results for the two crack models are illustrated in Figure \ref{fig14}, which plots the Von-Misses stress contour. It is observed that the high stress is concentrated in the crack tip area. The results agree with the theoretical analysis that singularities near crack tips generate a large stress around. In the following, parallel scalability is investigated by changing the subdomain solver and the number of processors. Based on the discussions above, the subsolvers ICC(8), ICC(9), and CC are considered for the two cases. The overlapping size for regular subdomains is 2, and crack tip subdomains is 6.

\begin{figure}[htb]
	\centering  
	\subfloat[ ] {\label{fig15a}
		\begin{minipage}[t]{0.5\textwidth}
			\centering       
			\includegraphics[scale=0.4]{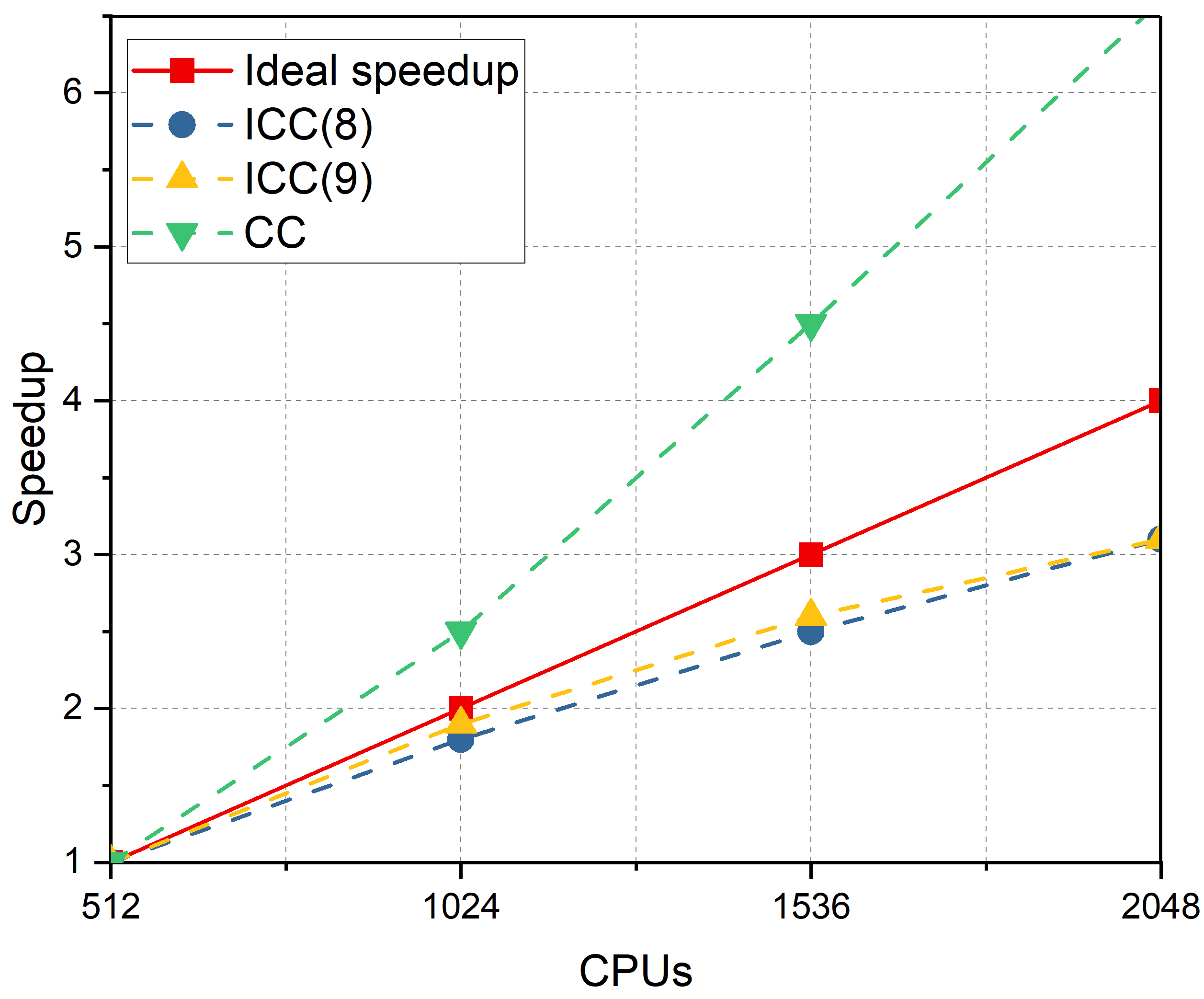}
		\end{minipage}%
	}
	\subfloat[ ]	{\label{fig15b}
		\begin{minipage}[t]{0.5\textwidth}
			\centering    
			\includegraphics[scale=0.4]{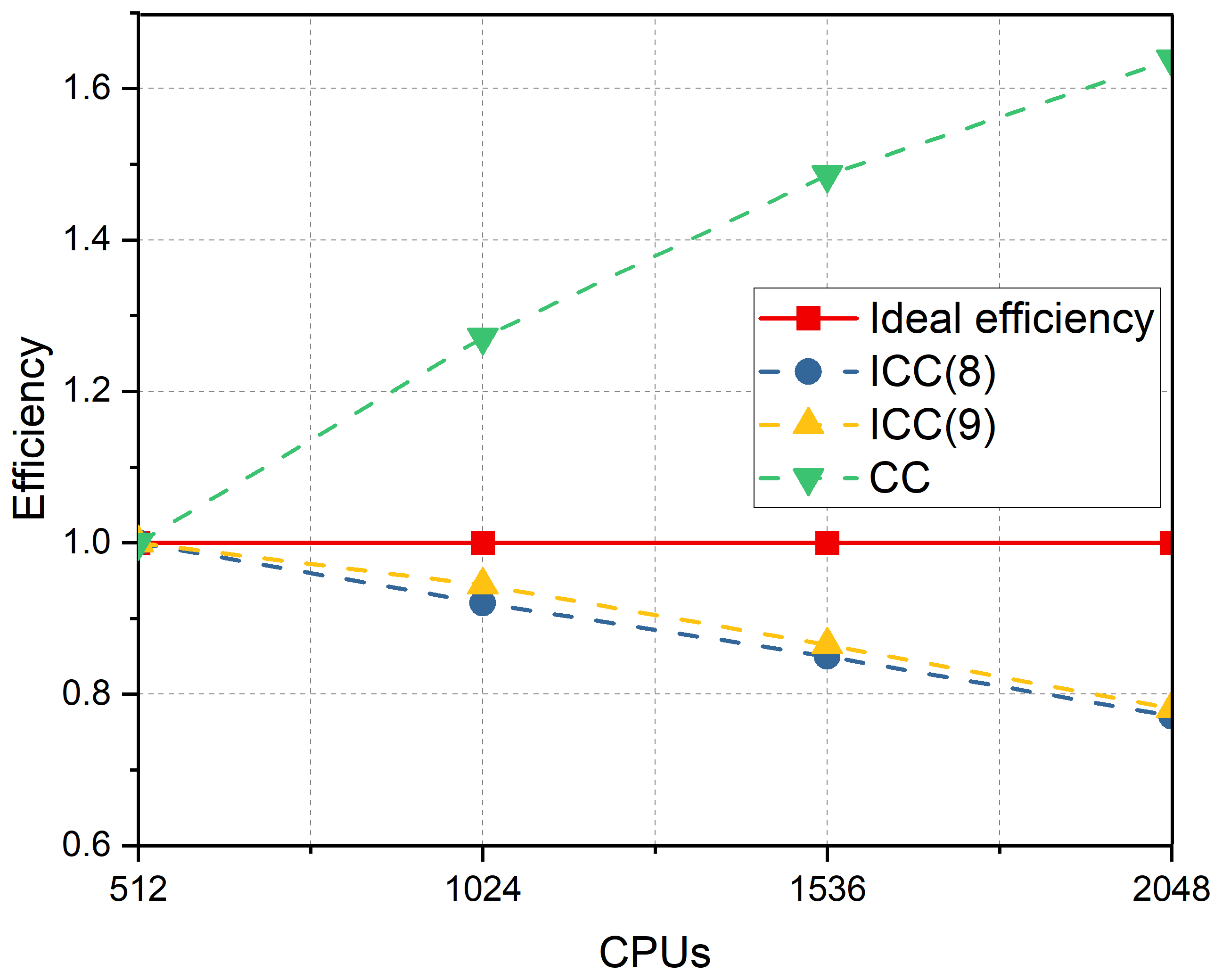}   
		\end{minipage}
	}
	\caption{Parallel scalability with respect to the number of processors for the branch crack model. (a) Speedup vs. number of processors; (b) efficiency vs. number of processors} 
	\label{fig15}  
\end{figure}

For the branch crack model, the mesh size is $5000\times5000$. After XFEM discretization, there were $5\times10^{7}$ regular DOFs and 15964 enriched DOFs. The number of processor cores varied from 512, 1024, 1536, to 2048. The numerical results are summarized in Table \ref{tab8}. The number of iterations and elapsed time decrease with an increase of the ICC fill-ins level. When the subdomain solver is set to CC, the iterations decrease to a minimum, whereas the solver time becomes much longer than that of ICC($l$). As for scalability with respect to the number of processors, the iterations increase continuously, and the elapsed time decreases significantly. In the last column for $T_{tip}/T_{reg}$, the subsolver time for crack tip subdomains occupies a relatively small percentage, and has a small impact on the scalability. The speedup and parallel efficiencies are displayed in Figure \ref{fig15}. The speedup with the subdomain solver ICC($l$) keeps increasing but is smaller than the ideal speedup, and the efficiency keeps decreasing with the increase of processors. The situation is different for the subdomain solver CC. The speedup and efficiency both exceed the ideal speedup and ideal efficiency while the rate of increase continues to decrease. This is because the complexity of ICC(\textit{l}) has a linear dependence on the number of processors, whereas for subsolver CC, the dependence is super-linear.

\renewcommand{\arraystretch}{1.2}
\begin{table}[h]
	\centering
	\caption{The scalability for the branch crack model ($np$ is the number of processors; ITER is the number of iterations; $T_{sol}$ is the elapsed time for the solver; Ideal is the ideal speedup; $T_{tip}/T_{reg}$ denotes the ratio of the subsolver time of crack tip subdomains to that of regular subdomains in percentage; the mesh is 5000$\times$5000; the overlapping size for regular subdomains is 2 and for crack tip subdomains is 6) \label{tab8}}%
	\begin{tabular}{|m{1.6cm}<{\centering}|m{1.6cm}<{\centering}|m{1.2cm}<{\centering}|m{1.2cm}<{\centering}|m{1.2cm}<{\centering}|m{1.2cm}<{\centering}|m{1.3cm}<{\centering}|m{1.2cm}<{\centering}|}
		\hline
		$np$                    &Subsolver &ITER &$T_{sol} (s)$ &Ideal & Speedup &Efficiency &$T_{tip}/T_{reg}$ \\ \hline
		\multirow{3}{*}{512} 	& ICC(8)	&1149	&64.09 		&1	&1.0 	&100.0\%		&1.81\%     \\ \cline{2-8} 
								& ICC(9)	&1083	&62.97 		&1	&1.0 	&100.0\%		&1.44\%     \\ \cline{2-8} 
								& CC		&704	&324.91 	&1	&1.0 	&100.0\%		&0.22\%     \\ \hline
		\multirow{3}{*}{1024} 	& ICC(8)	&1206	&34.82 		&2	&1.8 	&92.0\%			&3.25\%   		\\ \cline{2-8} 
								& ICC(9)	&1128	&33.34 		&2	&1.9 	&94.4\%			&2.98\%   		\\ \cline{2-8} 
								& CC		&739	&127.76 	&2	&2.5 	&127.2\%		&0.47\%     \\ \hline
		\multirow{3}{*}{1536} 	& ICC(8)	&1266	&25.14 		&3	&2.5 	&85.0\%			&4.25\%   		\\ \cline{2-8} 
								& ICC(9)	&1183	&24.26 		&3	&2.6 	&86.5\%			&4.08\%   	\\ \cline{2-8} 
								& CC		&785	&72.89 		&3	&4.5 	&148.6\%		&0.92\%   	  \\ \hline
		\multirow{3}{*}{2048}   & ICC(8)	&1299	&20.77 		&4	&3.1 	&77.1\%			&5.90\%   	\\ \cline{2-8} 
								& ICC(9)	&1233	&20.16 		&4	&3.1 	&78.1\%			&5.61\%   	\\ \cline{2-8} 
								& CC		&804	&49.57 		&4	&6.6 	&163.9\%		&1.73\%     \\ \hline
	\end{tabular}
\end{table}
\renewcommand{\arraystretch}{1}

\begin{figure}[htb]
	\centering  
	\subfloat[ ] {\label{fig16a}
		\begin{minipage}[t]{0.5\textwidth}
			\centering       
			\includegraphics[scale=0.4]{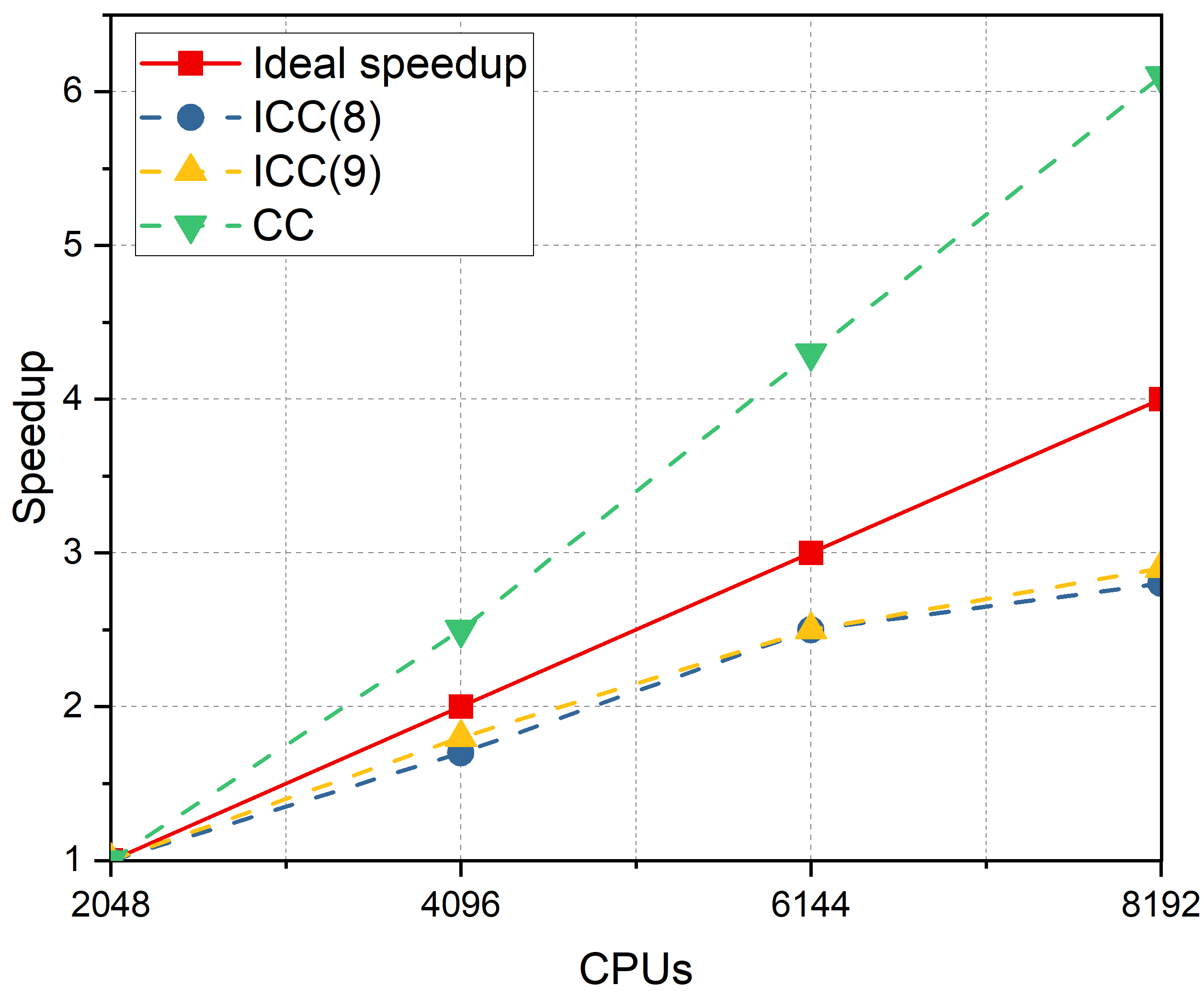}
		\end{minipage}%
	}
	\subfloat[ ]	{\label{fig16b}
		\begin{minipage}[t]{0.5\textwidth}
			\centering    
			\includegraphics[scale=0.4]{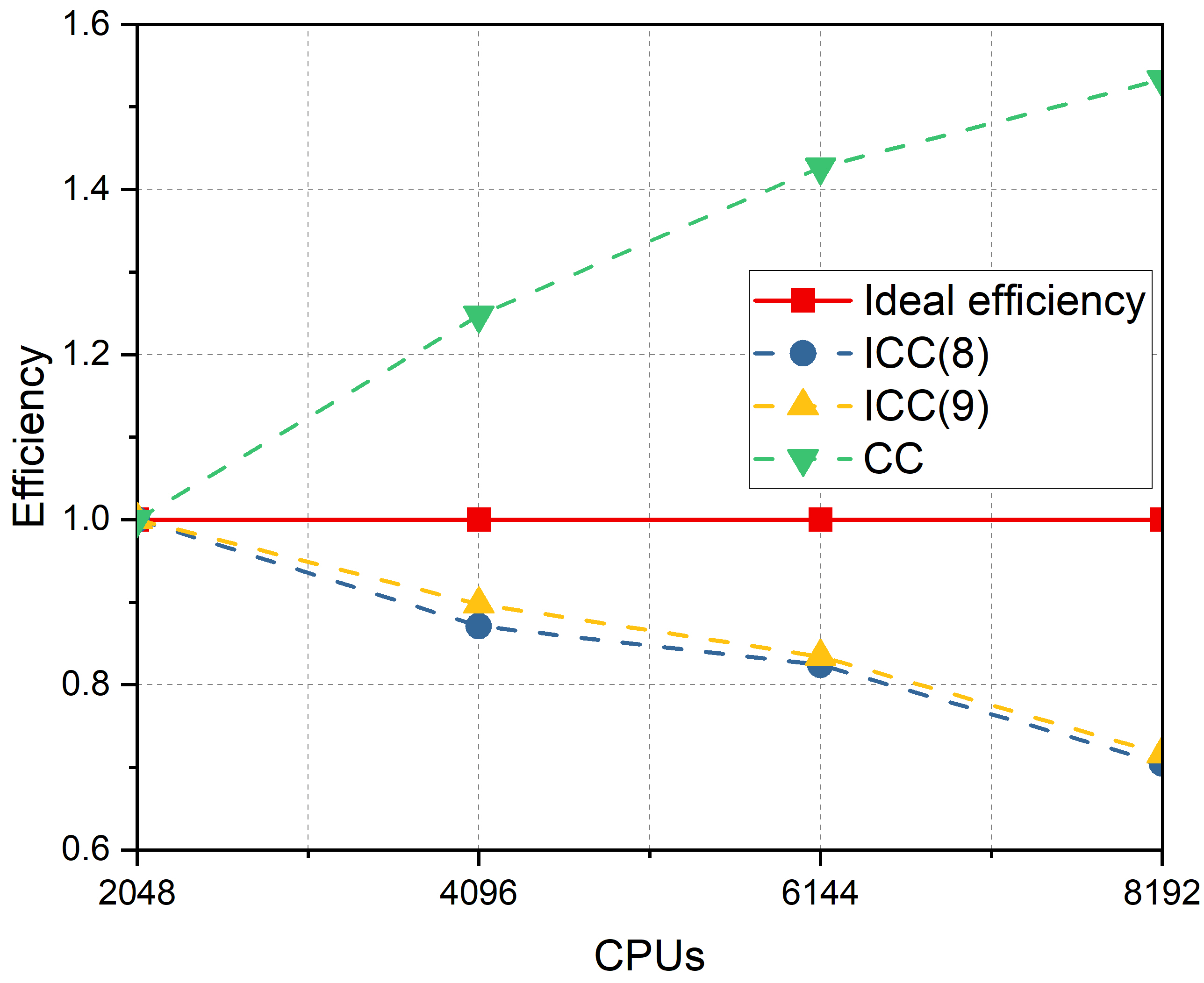}   
		\end{minipage}
	}
	\caption{Parallel scalability with respect to the number of processors for the 16-cracks model. (a) Speedup vs. number of processors; (b) efficiency vs. number of processors} 
	\label{fig16}  
\end{figure}

A more complex model is considered in Figure \ref{fig13b} to test the strong scalability of the proposed algorithm. The mesh is refined to $10000\times10000$, which has $2\times10^8$ regular DOFs and 123216 enriched DOFs after XFEM discretization. Based on the discussion so far, this large-scale problem is exceedingly difficult to solve because the condition number of the linear system is large (larger than $10^{20}$). Generally, the difficulty in solving a crack problem increases with mesh refinement and the number of cracks. If the domain has more cracks, the linear system includes more singularities, which makes the stiffness matrix more ill-conditioned. In this case, the subdomain solvers were ICC(8), ICC(9), and CC with different numbers of processors: 2048, 4096, 6144, and 8192. The numerical results are presented in Table \ref{tab9}. The results indicate that the basic principle of the subdomain solver and the scalability are the same as that of the branch crack model mentioned above. The difference is that the 16-cracks model requires more iterations and elapsed time to achieve convergence. In the last column for $T_{tip}/T_{reg}$, the subsolver time for the crack tip subproblem is considerable when the number of processors is 8192, as a result of the reduced parallel efficiency. The speedup and efficiency are displayed in Figure \ref{fig16}. Comparing Figure \ref{fig15} and Figure \ref{fig16}, the speedup and efficiency for the 16-cracks model are lower than those of the branch crack model with the same number of processors and ICC fill-ins level. 

\renewcommand{\arraystretch}{1.2}
\begin{table}[h]
	\centering
	\caption{The scalability for the 16-cracks model ($np$ is the number of processors; ITER is the number of iterations; $T_{sol}$ is elapsed time of solver; Ideal is the ideal speedup; $T_{tip}/T_{reg}$ denotes the ratio of subsolver time of crack tip subdomains to that of regular subdomains in percentage; the mesh is 10000$\times$10000; the overlapping size for regular subdomains is 2 and for crack tip subdomains is 6.) \label{tab9}}%
	\begin{tabular}{|m{1.6cm}<{\centering}|m{1.6cm}<{\centering}|m{1.2cm}<{\centering}|m{1.2cm}<{\centering}|m{1.2cm}<{\centering}|m{1.2cm}<{\centering}|m{1.3cm}<{\centering}|m{1.2cm}<{\centering}|}
		\hline
		$np$                    & Subsolver & ITER & $T_{sol} (s)$ & Ideal & Speedup & Efficiency &$T_{tip}/T_{reg}$\\ \hline
		\multirow{3}{*}{2048} 	& ICC(8)  	&1381	&77.34 		&1	&1.0 	&100.0\%	&2.67\%   \\ \cline{2-8} 
								& ICC(9) 	&1342	&77.83 		&1	&1.0 	&100.0\%	&2.59\%  	\\ \cline{2-8} 
								& CC	    &732	&340.73 	&1	&1.0 	&100.0\%	&0.45\%   	\\ \hline
		\multirow{3}{*}{4096} 	& ICC(8)   	&1438	&44.39 		&2	&1.7 	&87.1\%		&4.88\%	\\ \cline{2-8} 
								& ICC(9)	&1372	&43.35 		&2	&1.8 	&89.8\%		&4.49\%	\\ \cline{2-8} 
								& CC	    &781	&136.58 	&2	&2.5 	&124.7\%	&0.87\%   	\\ \hline
		\multirow{3}{*}{6144} 	& ICC(8)    &1496	&31.28 		&3	&2.5 	&82.4\%		&7.09\%	\\ \cline{2-8} 
								& ICC(9)	&1415	&31.09 		&3	&2.5 	&83.4\%		&6.58\%	\\ \cline{2-8} 
								& CC	    &821	&79.60 		&3	&4.3 	&142.7\%	&1.49\%   	\\ \hline
		\multirow{3}{*}{8192} 	& ICC(8)    &1529	&27.44 		&4	&2.8 	&70.5\%		&9.33\%	\\ \cline{2-8} 
								& ICC(9)	&1481	&27.16 		&4	&2.9 	&71.6\%		&9.22\%	\\ \cline{2-8} 
								& CC	    &863	&55.56 		&4	&6.1 	&153.3\%	&2.75\%   	\\ \hline
	\end{tabular}
\end{table}
\renewcommand{\arraystretch}{1}

The parallel analysis for the two crack models demonstrates that the proposed parallel domain decomposition preconditioner is scalable in terms of the number of processors. For a model with more cracks and refined mesh, the algorithm retains scalability. Despite the increase in the number of iterations and elapsed time to converge, the parallel efficiency remains at more than 70\%. Currently, the crack tip subproblem is solved on the part of processor cores, which decreases the parallel efficiency when the number of processors is large although the crack tip solver time is only a small percentage of that of regular subdomains. Such a restriction will be removed in future studies by solving the crack tip subproblem, employing all processors in parallel. 

\section{Conclusions}\label{sec5}
In this study, a parallel domain decomposition preconditioner was introduced to solve the elastic crack problem using corrected XFEM discretization. Ascribing the difficulty of the problem to the crack tip area, a special domain decomposition method was adopted, in which the regular subdomains include normal DOFs from standard FEM and Heaviside DOFs across cracks, and the crack tip subdomains include all types of DOFs within a radius $r_{tip}$ of the crack tips. An innovative algorithm based on the additive Schwarz method was proposed to speedup the iterations for solving the system of linear equations. In parallel computing, the algorithm solves the crack tip sub-problems first, then solves the regular sub-problems on all processors, and finally combines the two parts for the matrix-vector multiplication by the iterative solver. The effect of subdomain strategies, ICC fill-ins level, matrix reordering techniques, and overlapping size on the solver was further studied. The scalability and parallel efficiency of the algorithm were discussed using a branch crack model and 16-cracks model in a large number of processor cores. The numerical results demonstrate the effectiveness of the algorithm in terms of scalability and computation time. The parallel efficiency remained above 70\% with more than $2\times10^8$ DOFs. 

\section*{Acknowledgments}
This work was supported by the Special Project on High-performance Computing under the National Key R\&D Program (No. 2016YFB0200601) and Youth Program of National Natural Science Foundation of China (Project No. 11801543).

\subsection*{Conflict of interest}
The authors declare no potential conflict of interests.

\appendix
\section{eliminating linear dependency} \label{apda}
For the corrected XFEM, the branch enrichment functions are linearly dependent for a bilinear element. A description of a single crack tip was performed and presented in the work of T.P. Fries\cite{ref29}, which introduced a pair of parameters to remove the linear dependency. However, there are still issues to address, such as whether the choice of element and DOFs for the elimination are unique and which choice generates the minimum condition number for the linear system. In this section, we numerically discuss which equations to eliminate and how to eliminate them. The formulation of the branch enrichment functions in a quadrilateral element is
\begin{eqnarray}\label{eqa1}
	M_{ij}=\sum_{i=1}^{4}\sum_{j=1}^{4}N_i(x)[\phi_i^j(x)-\phi_i^j(x_i)]\mathcal{R}_i(x)
\end{eqnarray}
 
There are 16 local enrichment functions $M_{ij}$ in the element. For a two-dimensional problem, the number of corresponding DOFs is 32. In practice, there are many enrichment elements and enriched DOFs for each crack tip. These DOFs are distinguishable from other crack tips as a group. Of these 2 enrichment functions (4 enriched DOFs) have to be eliminated.

\begin{figure}[htb] 
	\centerline{\includegraphics[scale=0.15]{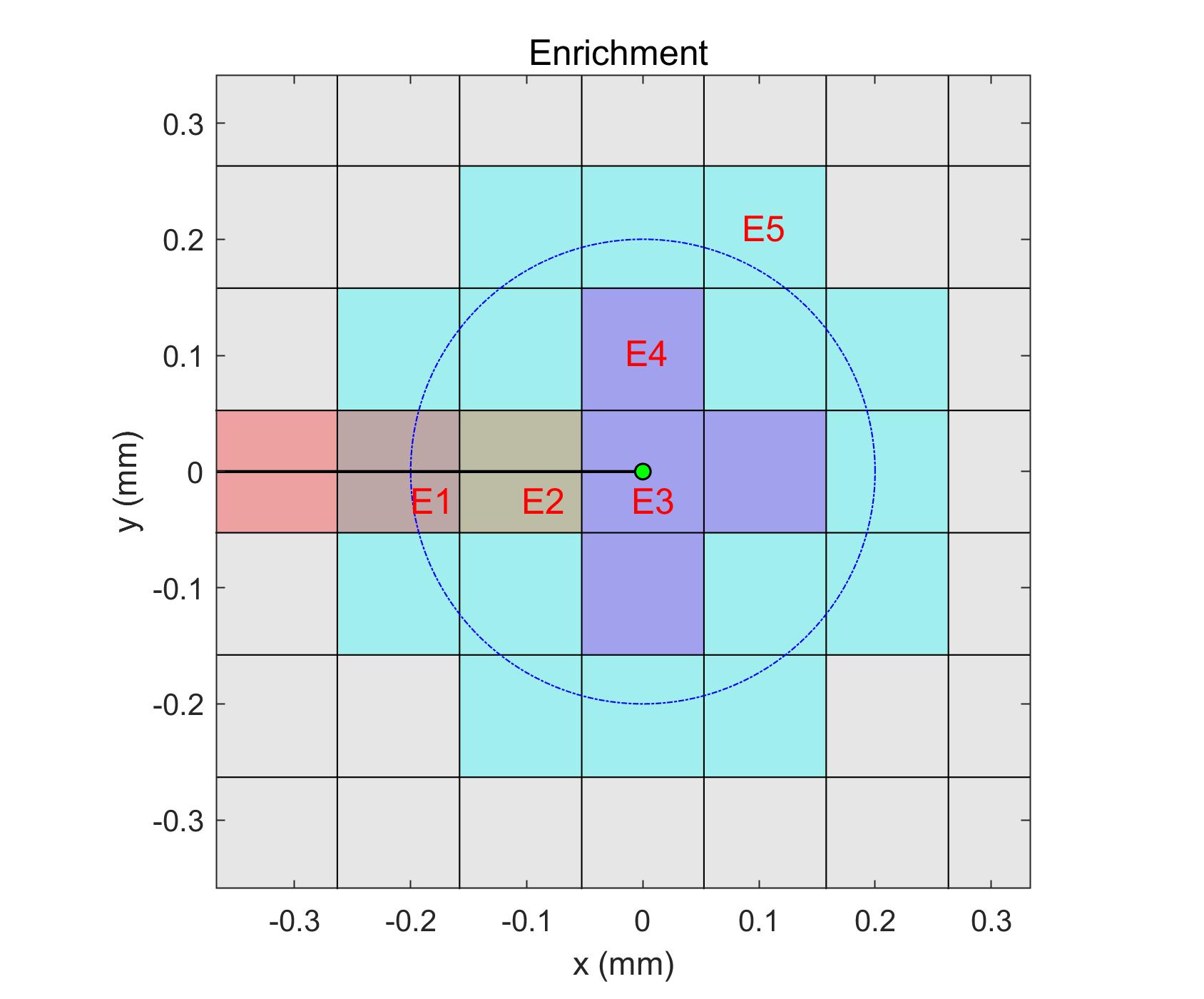}}
	\caption{elements and enrichments for a crack tip \label{figa1}}
\end{figure}

\begin{figure}[htp] 
	\centering
	\subfloat[E1] { \label{figa2a}
		\begin{minipage}[t]{0.33\textwidth}
			\centering       
			\includegraphics[scale=0.55]{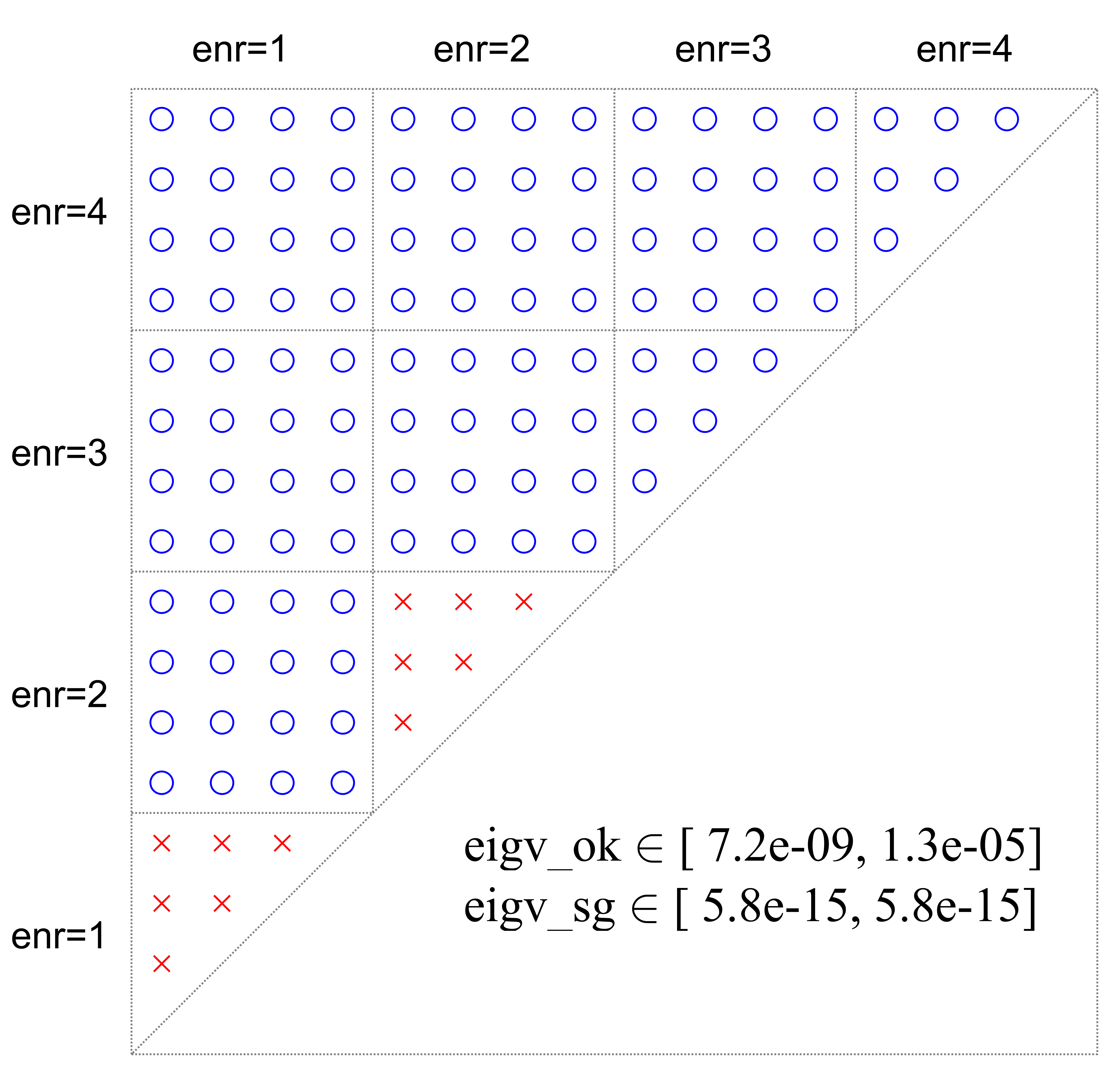}
		\end{minipage}%
	} 
	\subfloat[E2] {\label{figa2b}
		\begin{minipage}[t]{0.33\textwidth}
			\centering       
			\includegraphics[scale=0.55]{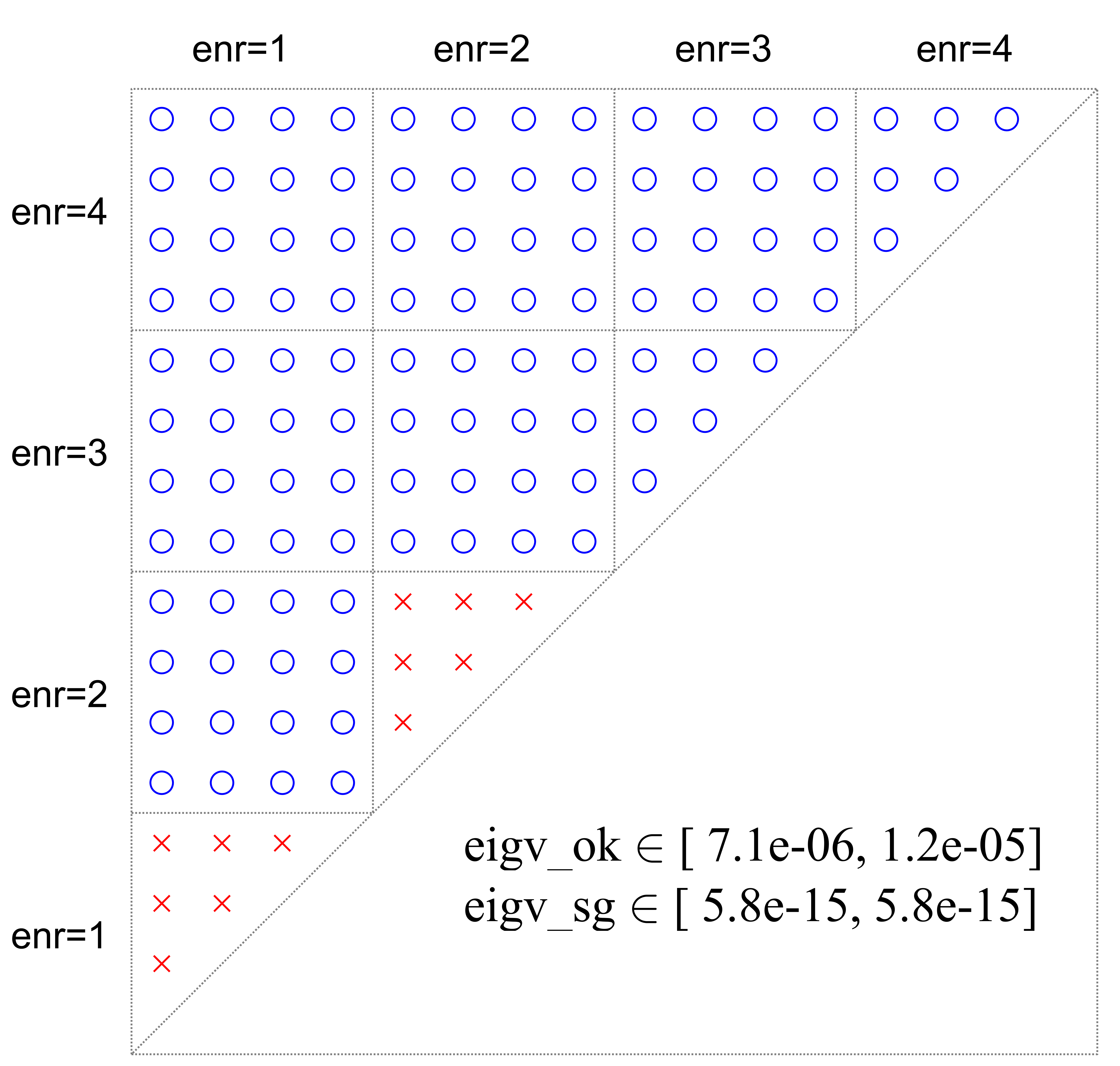}
		\end{minipage}%
	}
	\subfloat[E3] {\label{figa2c}
		\begin{minipage}[t]{0.33\textwidth}
			\centering       
			\includegraphics[scale=0.55]{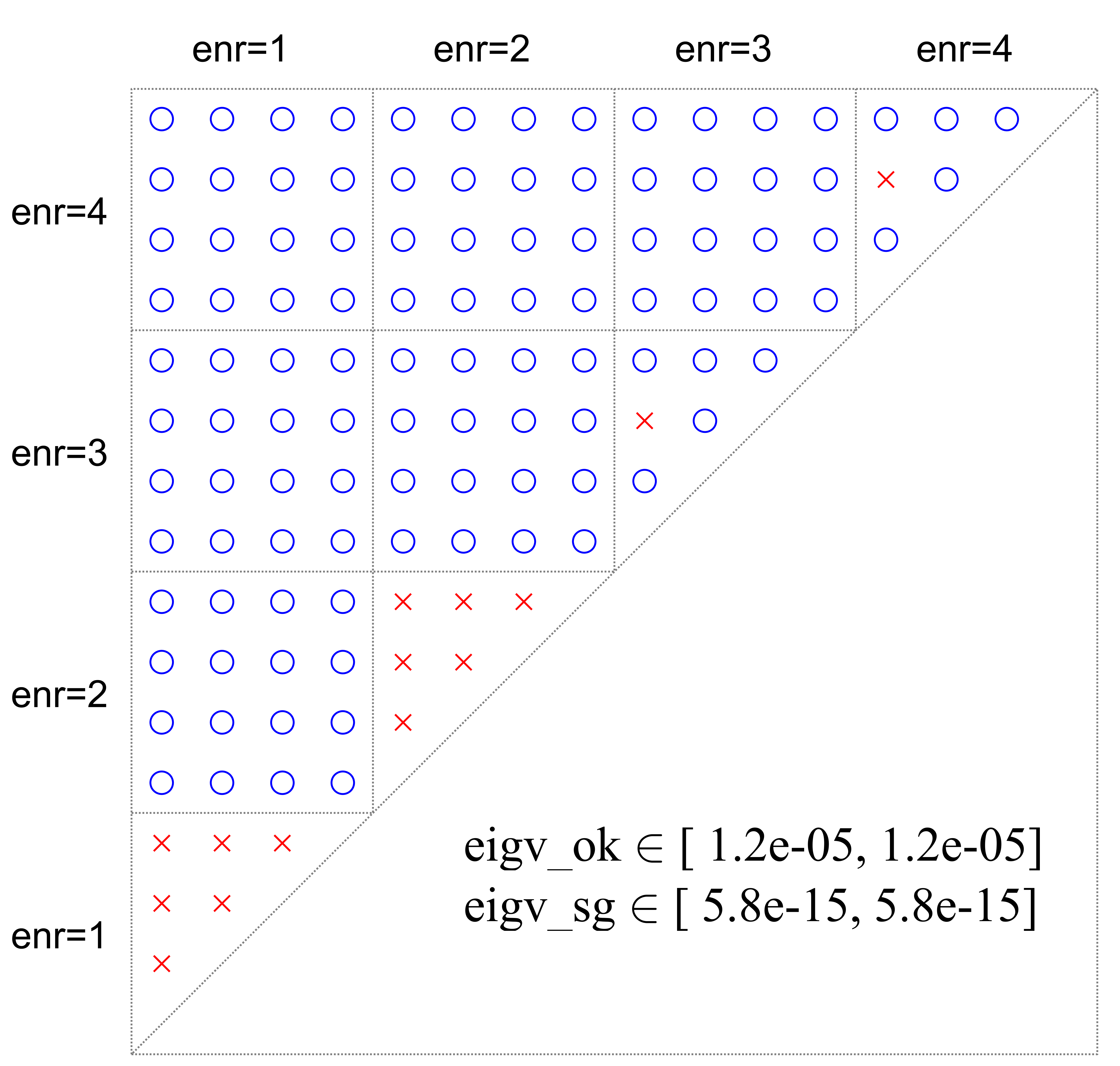}
		\end{minipage}%
	}
	
	\subfloat[E4] {\label{figa2d}
		\begin{minipage}[t]{0.33\textwidth}
			\centering       
			\includegraphics[scale=0.55]{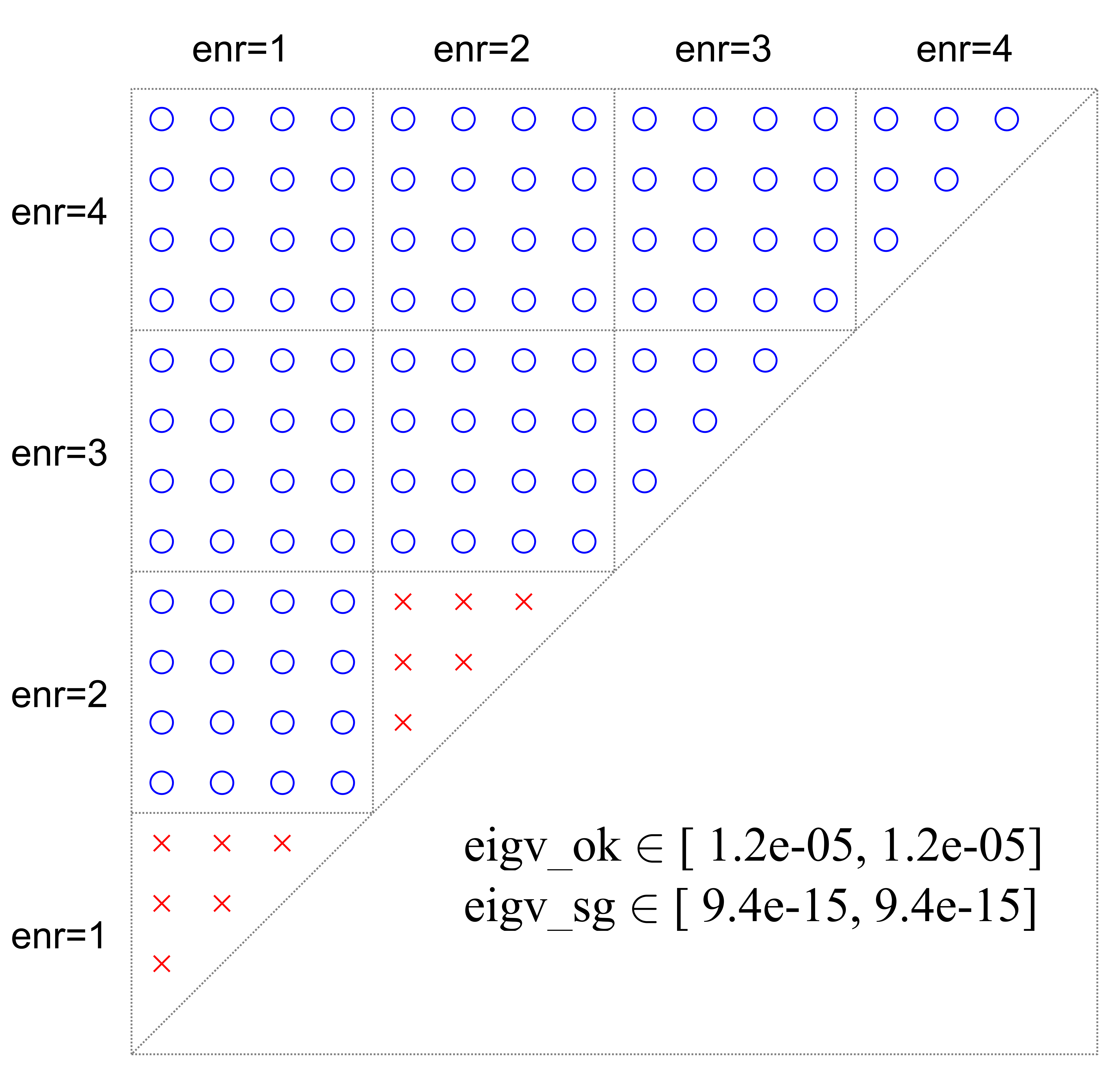}
		\end{minipage}%
	}
	\subfloat[E5] {\label{figa2e}
		\begin{minipage}[t]{0.33\textwidth}
			\centering       
			\includegraphics[scale=0.55]{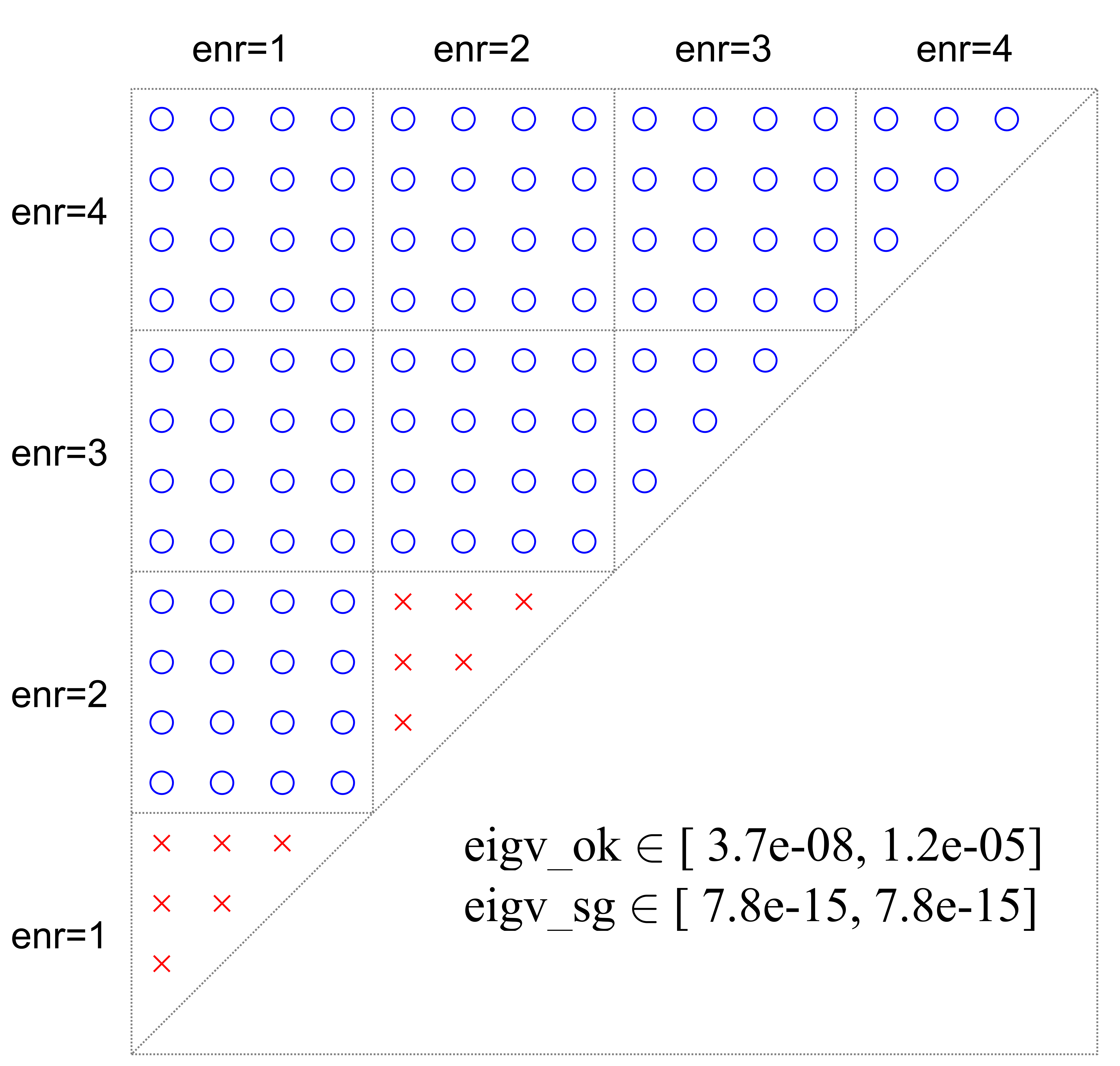}
		\end{minipage}%
	}
	\caption{The results of eliminating 2 of the 16 local enrichment functions for different elements, the '$\circ$' means the linear dependency has been removed and the '$\times$' means the linear dependency is preserved; the eigv\_ok is the range of minimum eigenvalues for good elimination and the eigv\_sg is the minimum eigenvalues range for the bad elimination. For example, eigv\_ok $\in[a, b]$ means that all these eliminations remove the linear dependency, and the minimum eigenvalues are larger than $a$ and smaller than $b$. \label{figa2}}
\end{figure}

In Figure \ref{figa1}, a simple case was employed to demonstrate the enrichment procedure for a crack tip. The element size is 0.1, and the enrichment radius $r_{tip}=0.185$, which results in only one layer of elements being completely in $r_{tip}$. Around the crack tip, 21 elements were enriched by the branch enrichment functions, and the number of enriched nodes was 32. There are five types of enrichment elements: E1 represents the blending element cut by a crack; E2 represents an element cut by a crack, and all nodes are within $r_{tip}$; E3 is the crack tip element; E4 is the enriched element without cracks inside; and E5 is the blending element without cracks inside. Two local enrichment functions need to be eliminated from these five types of elements, and we study whether the local enrichment functions are still linearly dependent after elimination by comparing the minimum eigenvalues. The threshold value is $1\times10^{-10}$, and if the minimum eigenvalue is larger than this value, the elimination to remove the linear dependency has been successful (called good elimination); otherwise, the linear dependency is preserved (called bad elimination). In Figure \ref{figa2}, all possible combinations, $C^2_{16}=120$, are displayed. eigv\_ok denotes the range of minimum eigenvalues for good elimination, and eigv\_sg denotes the minimum eigenvalue range for bad elimination, which preserves linear dependency. It is observed that the two enrichment functions based on the first or the second term in equation \ref{eq9}, completely preserve the linear dependency for E1, E2, E4, and E5. For the crack tip element E3, two additional combinations also maintain a linear dependency. Considering the range of minimum eigenvalues, the magnitude of the minimum eigenvalue for all bad eliminations is $10^{-15}$. For good elimination in E3 and E4, the magnitude of the minimum eigenvalue is $10^{-5}$. While for E1, E2, and E5, the magnitude of the minimum eigenvalues is $10^{-9}$, $10^{-6}$, and $10^{-8}$, respectively, but all are larger than $10^{-10}$. Therefore, eliminating two local equations in E3 or E4 is a good choice as the linear system after this elimination has the largest minimum eigenvalue. 

For a multi-crack model, two enrichment terms (four enriched DOFs) were eliminated for each crack tip in the linear system, as discussed above. In our code, we eliminated the 12th and 16th functions for the E4 element, which means that the last two terms in (\ref{eq9}) of the 4th node for an E4 element will be eliminated. All the results in this study were obtained after this elimination.

\bibliography{wileyNJD-AMA}

\clearpage

\end{document}